\documentclass[12pt]{iopart}%
\usepackage{iopams}
\usepackage{graphicx}
\usepackage{amssymb}
\usepackage{amstext}
\usepackage{amsfonts}%
\providecommand{\U}[1]{\protect\rule{.1in}{.1in}}

\newcommand{\sd}{\downarrow}
\newcommand{\su}{\uparrow}
\newcommand{\tp}{{ t^\prime}}
\begin{document}

\title{Manipulating magnetism by ultrafast control of the exchange interaction}
\author{J.H. Mentink$^{1}$}

\begin{abstract}
In recent years, the optical control of exchange interactions has emerged as an exciting new direction in the study of the ultrafast optical control of magnetic order. Here we review recent theoretical works on antiferromagnetic systems, devoted to i) simulating the ultrafast control of exchange interactions, ii) modeling the strongly nonequilibrium response of the magnetic order and iii) the relation with relevant experimental works developed in parallel. In addition to the excitation of spin precession, we discuss examples of rapid cooling and the control of ultrafast coherent longitudinal spin dynamics in response to femtosecond optically induced perturbations of exchange interactions. These elucidate the potential for exploiting the control of exchange interactions to find new scenarios for both faster and more energy-efficient manipulation of magnetism.
\end{abstract}

\address{$^1$Radboud University Nijmegen, Institute for Molecules and Materials,
Heijendaalseweg 135, 6525 AJ, Nijmegen, The Netherlands.} \ead{J.Mentink@science.ru.nl}

\section{Introduction}

Ordering of microscopic spins in magnetic materials originates from the exchange interaction $J_\text{ex}$, the strongest interaction in magnetism, which exceeds the strength of external magnetic fields by orders of magnitude. On a fundamental level, exchange interactions emerge from the repulsive Coulomb interactions between electrons and are most sensitive to electronic perturbations. This fact implies intriguing possibilities for the ultrafast control of magnetism by femtosecond laser pulses, which is a very active research field initiated two decades ago with the ground breaking discovery of sub-picosecond demagnetization of ferromagnetic Ni by a 60 femtosecond laser pulse \cite{beaurepaire1996} and the observation of laser-induced ferromagnetic \cite{vankampen2002} and antiferromagnetic resonance \cite{kimel2005}, followed by the observation of all-optical switching in ferrimagnetic GdFeCo alloys \cite{stanciu2007,vahaplar2009,kirilyuk2010,ostler2012} and subsequently the highly intriguing observation of distinct dynamics between exchange coupled spins in different magnetic sublattices \cite{radu2011,mathias2012,radu2015}. Moreover, a further stimulus to the field was given by the demonstration of helicity dependent all-optical switching in ferromagnetic multilayers \cite{lambert2014}, which are materials of great interest for magnetic data storage and, very recently, by the demonstration of all-optical magnetic recording in transparent ferrimagnetic oxides \cite{stupakiewicz2017}, enabling magnetic recording that is both ultrafast and takes place at unprecedentedly low heat load. 

\begin{figure}[t]
\centering{\includegraphics[width=\columnwidth]{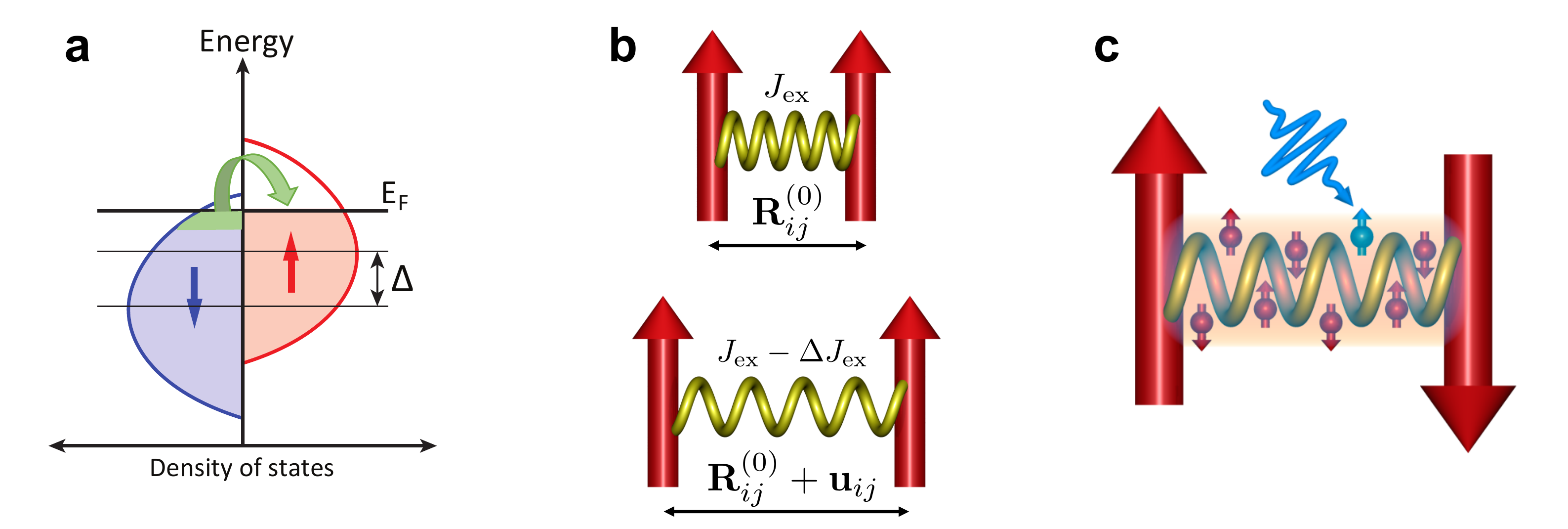}}
\caption{
{\bf Illustrated examples of the effect of laser excitation on exchange effects in magnetic systems}. (a) Collapse of the exchange splitting in itinerant ferromagnets as a result of the redistribution of laser-excited electrons between the majority and minority band (green arrow). (b) optical excitation of phonons change the positions of the atomic nuclei $R_{ij}$, leading to a perturbation of the exchange interaction (yellow spirals) between spins localized around the different nuclei (red arrows). (c) femtosecond laser excitation (blue pulse) of electrons (small spheres with arrows) can change both the band structure and the electronic distribution. This yields perturbations to the exchange interaction (glowing yellow spiral) between spins localized at different sites (big red arrows). Case (c) is the focus of this review.
\label{f:intro}}
\end{figure}

All the above experiments can be understood by accounting for (a combination of) laser-induced heating, generation of effective opto-magnetic fields and/or optical perturbations to the magnetic anisotropy, but do not directly provide indications for time-dependent exchange interactions. Interestingly, however, considerable experimental evidence has been has been presented as well for dynamical exchange effects, including a collapse of the exchange splitting in Ni, Gd and Co metals \cite{rhie2003,carley2012,frietsch2015,eiche2017} (illustrated in Fig.~\ref{f:intro}a), modulation of $J_\text{ex}$ by excitation of an optical phonon at the Gd metal surface \cite{melnikov2003} (Fig.~\ref{f:intro}b), laser-induced heating across the antiferromagnetic-ferromagnetic transition in FeRh \cite{ju2004,thiele2004} and, more recently, detection of the dynamics of the exchange energy \cite{subkhangulov2014}, as well as triggering both coherent macroscopic spin precession \cite{mikhaylovskiy2015} and longitudinal oscillation of the order parameter in magnetic insulators \cite{bossini2016} by optical perturbations of $J_\text{ex}$ (Fig.~\ref{f:intro}c). Furthermore, the ability to control the exchange interaction by time-dependent electric fields has intrigued researchers in several other areas of physics, including quantum computing based on semiconductor quantum dots \cite{loss1998,shahbazyan2000,piermarocchi2002},  
ultracold atoms \cite{duan2003,trotzky2008,chen2011}, strongly correlated materials \cite{wall2009,forst2011,li2013} and semiconductors doped with impurity spins \cite{nagaev1988,kane1999,piermarocchi2004,fernandez-rossier2004,wang2007,matsubara2015}. 

Despite this significant amount of studies, the problem of understanding and modeling how magnetism can be manipulated by ultrafast control of exchange interactions is still far from being solved. In particular, while considerable progress has been made for describing optical control of $J_\text{ex}$ in one and two spin systems \cite{loss1998,shahbazyan2000,piermarocchi2002,duan2003} the generalization to extended magnetically ordered systems is non-trivial as it requires to understand how laser pulses influence both the band structure and the electronic correlations. This is a highly challenging problem, since it implies the solution of a strongly time-dependent quantum many-body problem of an extended system. Moreover, when exchange interactions are perturbed on time scales much shorter than the equilibration of the magnetic system, the latter can be brought in a strongly nonequilibrium state which cannot be treated within a conventional thermodynamical approach. Even in the semi-classical regime, where spins can be treated as classical vectors, simulating the time-dependent response of the macroscopic magnetic order to ultrafast modifications of atomic-scale exchange interactions defines a challenging multi-scale problem.

Early works addressing both the optical control of exchange interactions and the response of magnetic order focused on ferromagnets, in particular ferromagnetic semiconductors \cite{fernandez-rossier2004}. However, in ferromagnetic systems the excitation of spin dynamics requires a change of the total angular momentum. This makes it difficult to induce fast dynamics by modifying $J_\text{ex}$. Antiferromagnetic (AFM) systems do not suffer from this bottleneck and therefore provide novel opportunities to manipulate the dynamics of magnetic order by ultrafast control of $J_\text{ex}$. This review article provides an overview of the recent developments for modeling such antiferromagnetic systems \cite{mentink2014,mentink2015,mikhaylovskiy2015,hellsvik2016,bossini2016} and is organized as follows. First, we introduce various analytical methods that enable the definition of exchange interactions under electronic nonequilibrium conditions and outline the computational methods used to evaluate the resulting formulas. Second, results on the ultrafast control of $J_\text{ex}$ are discussed using the single-band Mott-Hubbard insulator at half-filling as a model system, which allows to study both resonant electronic excitations as well as non-resonant periodic driving. Third, the manipulation of magnetic order by the ultrafast control of exchange interactions is discussed, focusing on four examples: a) the excitation of macroscopic spin precession in canted antiferromagnets, b) cooling of antiferromagnetically ordered classical spins c) excitation of coherent longitudinal oscillations of the AFM order parameter, and d) effective time reversal in quantum spin chains. Finally, we draw conclusions and discuss several directions for further research.

\section{Methods of computing time-dependent exchange interactions}\label{s:methods}
Most of the calculations on the ultrafast control of $J_\text{ex}$ reported here are based on calculations using the paradigm single band Hubbard model. The advantage of using the single band Hubbard model is that $J_\text{ex}$ in equilibrium is very well understood and serves as minimal model for describing exchange interactions in magnetic oxides. At the same time, recently established computational techniques can be exploited to study the nonequilibrium electron dynamics for extended systems, enabling the evaluation of $J_\text{ex}$ out of equilibrium. Below we start by briefly introducing the Hubbard model and subsequently outline the methods used to solve it out of equilibrium. Finally, we introduce three distinct methods to evaluate $J_\mathrm{ex}$ under electronic nonequilibrium conditions.

\subsection{Hubbard model}
The Hamiltonian of the Hubbard model is given by
\begin{equation}
\label{repulsive hubbard}
H
=
-t_0\sum_{\langle ij \rangle \sigma}
c_{i\sigma}^\dagger
c_{j\sigma}
+
U
\sum_{j}
n_{j\uparrow}n_{j\downarrow}
\end{equation}
Here $c_{i\sigma}^\dagger$ creates an electron at site $i$ with spin $\sigma=\su,\sd$, $t_0$ is the hopping between nearest-neighbor sites and $U$ the repulsive on-site interaction $U$. For half-filling and at $U/t_0\gg1$ this model describes a Mott-insulator with one electron per site. The AFM exchange coupling between the spin degrees of freedom follows from the well-known kinetic exchange mechanism \cite{anderson1959}, where the system gains energy by virtual hoppings to adjacent sites. Due to the Pauli principle, such hoppings are only possible when adjacent sites have opposite spin. A simple perturbative calculation in $t_0$ shows that the spin degrees of freedom are described by an AFM Heisenberg Hamiltonian $H_\text{ex}=J_\text{ex}\sum_{ij}\mathbf{S}_i\cdot\mathbf{S}_j$, where $J_\mathrm{ex}=2t_0^2/U$ is the exchange interaction (see also Fig.~\ref{f:floq}a), which has been derived more rigorously based on a canonical transformation technique \cite{harris1967,chao1977,takahashi1977,macdonald1988}. 

For the evaluation of $J_\mathrm{ex}$ below, it is convenient to include also a homogenous static magnetic field described by
\begin{equation}
H_Z=B_x
\sum_j
S_{jx}.
\end{equation}
where the spin $S_{j\alpha}=\frac{1}{2} \sum_{\sigma\sigma'} c_{j\sigma}^\dagger (\hat \sigma_\alpha)_{\sigma\sigma'} c_{j\sigma'}$ is coupled to a homogeneous magnetic field $B_x$ along the $x$ axis ($\alpha=x,y,z$; $\hat \sigma_\alpha$ denote the Pauli matrices). Non-equilibrium dynamics due to time-dependent electric fields $\mathbf{E}(t)$ can be most conveniently incorporated by including a Peierls phase \cite{peierls1933,luttinger1951} to the hopping: 
\begin{equation}
t_{ij}(t)=t_0\exp\left[{\mathrm{i}e\mathbf{A}(t)\cdot(\mathbf{R}_i-\mathbf{R}_j)/\hbar}\right],
\end{equation}
where $\mathbf{A}(t)=-\partial_t\mathbf{E}(t)$ is the homogenous vector potential. Physically, this is equivalent to using a gauge in which the electric field is included by adding a scalar potential at each site, $e\phi_i(t)\sim\mathbf{R}_i\cdot\mathbf{E}(t)$. 

To solve the Hubbard model in the presence of time-dependent electric fields, we outline two complementary methods used to obtain the results discussed in this review. First, the nonequilibrium extension of the Dynamical Mean Field Theory (DMFT), which enables the investigation of magnetically ordered systems directly in the thermodynamic limit. Second, direct simulation of the time-dependent Schr\"odinger equation using exact diagonalization for finite size one-dimensional systems.

The implementation of nonequilibrium DMFT was reviewed in detail in \cite{aoki2014}. In short, within DMFT \cite{georges1996} the problem of interacting electrons on a lattice is mapped onto the solution of an effective single-site impurity coupled to a non-interacting bath, that is determined self-consistently. This results in a mean field theory for the spatial degrees of freedom while keeping temporal correlations and was shown to be the exact solution of the Hubbard model in the limit of infinite dimensions \cite{metzner1989}. For the nonequilibrium case, the model can be solved using the perturbative hybridization expansion (non-crossing approximation \cite{eckstein2010nca}), which at large $U$ shows good agreement with more accurate impurity solvers. Further, to solve the Hubbard model in the presence of a transverse magnetic field, spin-flip terms were included in the solution of the impurity model. The detailed implementation for the hypercubic lattice is described in \cite{mentink2014}.

Exact diagonalization solves directly the Schr\"odinger equation with the time-dependent Hamiltonian $H(t)$ from a given initial state $|\psi_0\rangle$. For systems of a few sites, evaluation of the time propagator $\exp{(-\mathrm{i}H(t)/\hbar})$ can be done directly by numerically diagonalizing $H(t)$. For larger systems, a more efficient scheme is required and the Krylov technique was used \cite{hochbruck1997}. In both cases, this is combined with a commutator-free exponential time-propagation scheme \cite{alvermann2011}. Being related to the Magnus expansion, it preserves unitarity and yields a high-order accurate time integration of the Schr\"odinger equation.

\subsection{Nonequilbrium exchange in canted antiferromagnets}\label{s:jexcanted}
\begin{figure}[t]
\centering{\includegraphics[width=0.5\columnwidth]{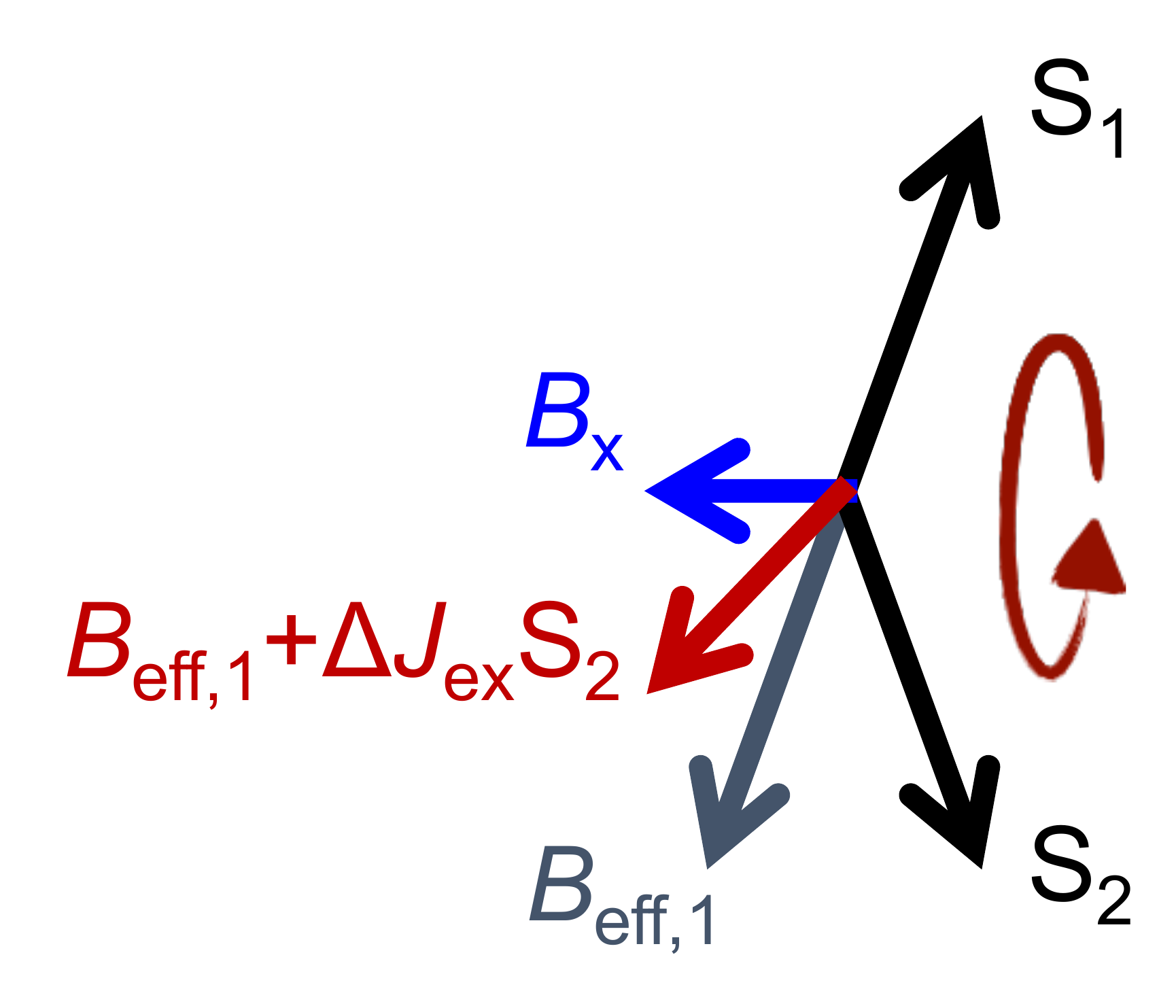}}
\caption{
{\bf Evaluation of the nonequilibrium $J_\text{ex}(t)$ in a canted antiferromagnet}. In equilibrium, the effective field $B_{\text{eff},1}$ (grey arrow) is antiparallel to the sublattice magnetization $\mathbf{S}_1$ (upper black arrow). A modification of the exchange interaction ($\Delta J_\text{ex}$) would rotate the effective field (red arrow) with respect to $\mathbf{S}_1$ causing the excitation of a spin resonance. In turn, $\Delta J_\text{ex}(t)$ can be inferred from the observed spin precession. 
\label{f:canted}}
\end{figure}

A simple instrumental way to evaluate a nonequilibrium exchange interaction is by studying a two-sublattice antiferromagnet canted by an external magnetic field \cite{mentink2014}. In the regime where a rigid macrospin model would be valid, the dynamics of the sublattice macrospin is governed by the Landau-Lifshitz equation 
\begin{equation}\label{e:ll}
\hbar\partial_t\langle\mathbf{S}_1\rangle = - \langle\mathbf{S}_1\rangle\times\left(B_x\mathbf{e}_x-2J_\mathrm{ex}\langle\mathbf{S}_2\rangle\right),
\end{equation}
where $\langle\mathbf{S}_{1,2}\rangle$ is the macrospin on sublattice 1,2; $J_\mathrm{ex}$ is the effective exchange interaction and $B_x$ is the strength of the transverse magnetic field. With $\langle\mathbf{S}_{1,2}(t)\rangle$ given from the solution of the full electronic model, the effective exchange interactions can be inferred from inverting Eq.~(\ref{e:ll}). Using N\'eel symmetry, $\langle S_{1y,z}\rangle = -\langle S_{2y,z}\rangle$, $\langle S_{1x}\rangle=+\langle S_{2x}\rangle$ one obtains
\begin{equation}\label{e:jexcanted}
J_\mathrm{ex}(t)=-\frac{B_x}{4\langle S_{1x}\rangle} - \frac{1}{4\langle S_{1x}\rangle}\frac{\hbar\partial_t\langle S_{1y}\rangle}{\langle S_{1z}\rangle}.
\end{equation}
This result can be interpreted as follows. The first term on the right gives the equilibrium value of $J_\mathrm{ex}$ which is determined by the canting induced by $B_x$. As illustrated in Fig.~\ref{f:canted}, in this case the effective field $\mathbf{B}_1=B_x\mathbf{e}_x-2J_\mathrm{ex}\langle\mathbf{S}_2\rangle$ is parallel to $\langle\mathbf{S}_1\rangle$ and no dynamics occurs. If however, $J_\mathrm{ex}$ is suddenly perturbed by an amount $\Delta J_\mathrm{ex}$, the effective field is no longer collinear to $\langle\mathbf{S}_1\rangle$ and a precession is triggered, leading to the appearance of the second term in Eq.~(\ref{e:jexcanted}). While this interpretation may seem conceptually attractive, we stress that the validity of the instantanous approximation used in the derivation of Eq.~(\ref{e:ll}) is a fundamental question which in general has not been solved (see also the discussion in \cite{mentink2014}). For the purpose of this review, we regard Eq.~(\ref{e:jexcanted}) as the best estimate of an instantaneous $J_\mathrm{ex}$ that is in accordance with an observed macrospin dynamics. 

\subsection{General formulas for nonequilibrium exchange}\label{s:jexgeneral}
Instead of explicitly simulating the full electron dynamics in the canted geometry, it is also possible to derive explicit formulas for the response to small rotations of spins. In equilibrium, this approach was introduced by Lichtenstein, Katsnelson and coworkers \cite{lichtenstein1984,lichtenstein1985,lichtenstein1987,katsnelson2000}. For two spin moments at site $i$ and $j$ rotated by a small angle $\pm \theta/2$ the relative energy change can be written as $\delta E_{ij}\approx \frac{1}{2}J_{ij}\theta^2$. In the single band Hubbard model the pair-interactions than become \cite{katsnelson2000}
\begin{equation}
J_{ij}=-\mathrm{Tr}_E \left(\mathit{\Sigma}_i^s(E)G^\su_{ij}(E)\mathit{\Sigma}^s_j(E)G^\sd_{ji}(E) \right)
\end{equation}
Here $\mathit{\Sigma}_i^s(E)=\left(\mathit{\Sigma}_i^\su(E)-\mathit{E}_i^\sd(\omega)\right)/2$ is the spin-dependent part of the self-energy which is taken only locally at site $i$. $G^\sigma_{ij}(E)$ is the spin-dependent single-particle Green's function and the trace is over all electron energies $E$. It is useful to realize that already this equilibrium result indicates a profound difference between the exchange splitting observed in photo-emission experiments and the intersite exchange interaction $J_{ij}$ responsible for ordering of spin moments. While in simplest approximation the exchange splitting is determined $\mathit{\Sigma}^s_i(0)$ (Stoner model), the $J_{ij}$'s are also dependent on the Green's function and hence even in equilibrium it is not possible to directly extract intersite exchange interactions from photo-emission data. 

The concept of rotating local moments can be generalized also to the time-dependent case \cite{secchi2013}. In this case, an effective spin action is defined in terms of time dependent rotations of the spin quantization axes $\mathbf{e}_i(t)$, as described by Holstein-Primakoff bosons $\xi_i(t)$. Starting from the electronic partition function as a path integral over fermionic fields $\phi$, one introduces rotated fermion fields $\psi$ and then expands the action to second order in $\xi$. The rotated fermionic fields are integrated out, which leads to spin action with an interaction term of the form $\mathcal{S}_\text{spin}[\xi^*\!,\xi]=\sum_{ij}\int\!dt\!\int\!d\tp\,\xi_{i}^*(t)A_{ij}(t,t')\xi_{j}(\tp)$. The coupling $A_{ij}(t,\tp)$ between spin rotations at different times and different sites $i\neq j$ becomes
\begin{eqnarray}
A_{ij}(t,\tp) &=& \frac{1}{4}\left[R^\sd_{ij}(t,\tp)R^\su_{ji}(\tp,t) + S_{ij}^\sd(t,\tp)S_{ji}^\su(\tp,t) \right.\nonumber \\
&& \quad\left.- T^\sd_{ij}(t,\tp)G^{\su}_{ji}(\tp,t) - G^{\sd}_{ij}(\tp,t) T^\su_{ji}(\tp,t)\right],
\label{jex:A}
\end{eqnarray}
where $T^\sigma_{ij}(t,\tp)=\mathit{\Sigma}^\sigma_{ij}(t,\tp)+\left[ \mathit{\Sigma} \star G \star \mathit{\Sigma} \right]^\sigma_{ij}(t,\tp)$, $R^\sigma_{ij}(t,\tp)=\left[ G \star \mathit{\Sigma} \right]^\sigma_{ij}(t,\tp)$, $S^\sigma_{ij}(t,\tp)=\left[ \mathit{\Sigma} \star G \right]^\sigma_{ij}(t,\tp)$, and $\star$ denotes the convolution. In general, these expressions include retardation (memory) effects, as expressed by couplings $A_{ij}(t,\tp)$ depending on two time variables. This can be mapped onto an instantaneous exchange coupling when the rotations of the spin quantization axes are much slower than the electron dynamics, and, in particular, slower than time-dependent fluctuations of the local magnetic moments themselves. Averaging over the fast dynamics than gives
\begin{equation}
\label{jex:integral}
\overline{A_{ij}}(t)=\,\mathrm{Im}\! \int_0^\infty \!\!\!\!ds \,A^\mathrm{ret}_{ij}(t,t-s).
\end{equation}
Still, $\overline{A_{ij}}(t)$ contains not only the exchange interactions, but also the time-averaged reduction of the local spin by fluctuations. In the absence of symmetry breaking, the bare exchange interactions can be defined as
\begin{equation}\label{jex:bare}
J_{ij}(t)=\frac{\overline{A_{ij}}(t)}{\langle\mathbf{S}_i(t)\cdot\mathbf{S}_j(t)\rangle},
\end{equation}
where $\langle\mathbf{S}_i(t)\cdot\mathbf{S}_j(t)\rangle$ is the equal-time spin correlation function. To evaluate these formulas, one has to compute the Greens functions and self-energies. A practical advantage of the general formulas is that they enable evaluation of exchange interactions in the collinear state, for which evaluation of the electronic dynamics is generally simpler. The implementation of Eq.(\ref{jex:A})-(\ref{jex:bare}) based on exact diagonalization is discussed in \cite{mikhaylovskiy2015}, while the implementation within nonequilbrium DMFT is detailed in \cite{mentink2015}, where $\langle\mathbf{S}_i(t)\cdot\mathbf{S}_j(t)\rangle$ in Eq.~(\ref{jex:bare}) is replaced by $\langle S_{iz}(t)\rangle\langle S_{jz}(t)\rangle$. 

\subsection{Floquet theory of nonequilibrium exchange}\label{s:jexfloquet}
\begin{figure}[t]
\includegraphics[width=\columnwidth]{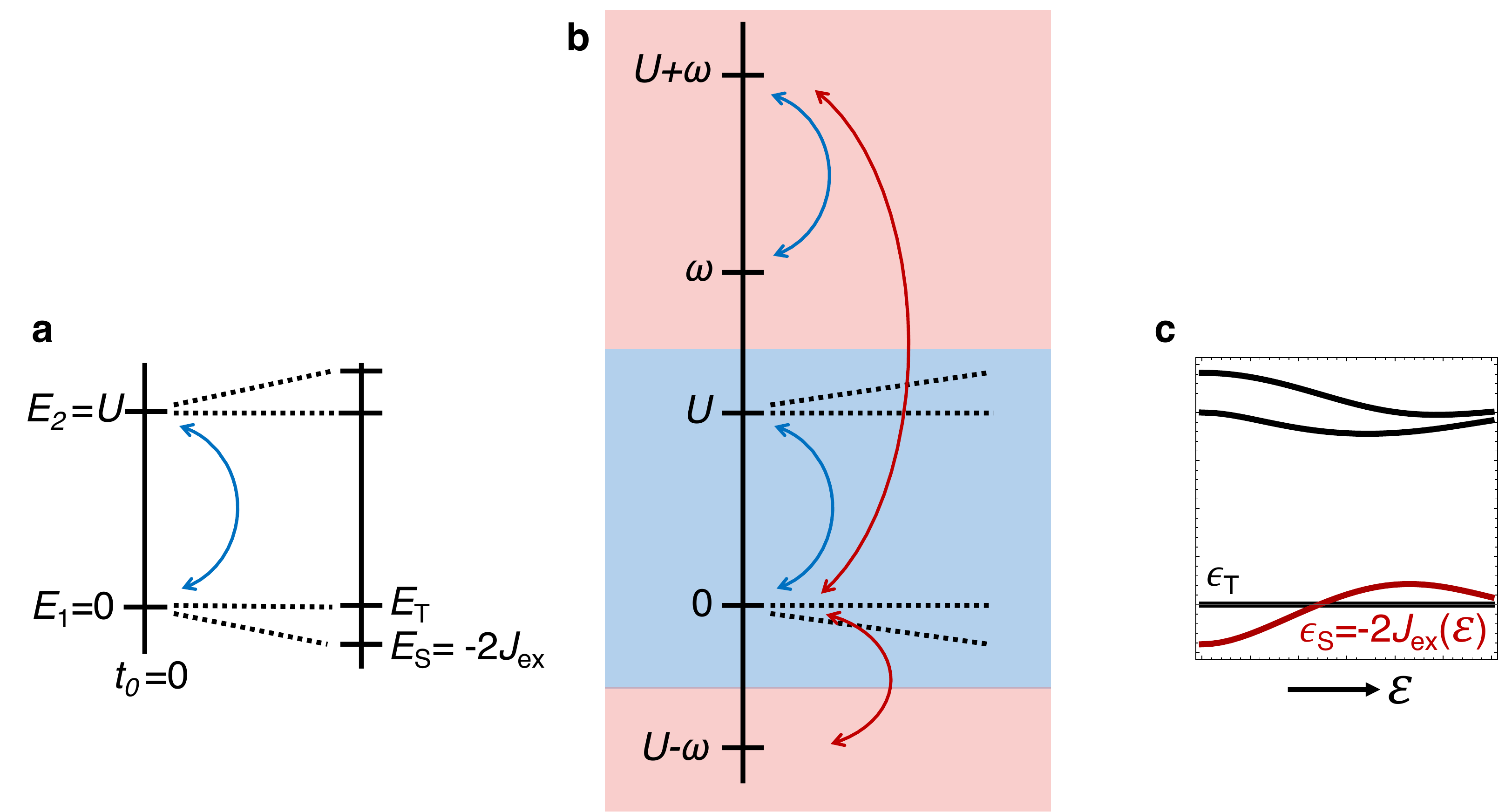}
\caption{
{\bf Floquet theory of nonequilibrium $J_\text{ex}$ in a two-site Hubbard model}. (a) In the atomic limit, there are four states. Two of them ($|\!\!\uparrow,\downarrow\rangle$ and $|\!\downarrow,\uparrow\rangle$) are singly occupied sites at $E_1=0$ while the other two states involve a doubly occupied and empty site at energy $E_2=U$ ($|\!\!\uparrow\!\downarrow,0\rangle$ and $|0,\uparrow\!\downarrow\rangle$). Due to virtual hoppings between these sets of states (blue arrows, the degeneracy is lifted and the lowest states become singlet and triplet states at energies $E_S=-2J_\text{ex}$ and $E_T=0$, respectively. (b) In the presence of a time-periodic field, there are also virtual hoppings induced by absorption and emission of photons (red arrows), coupling different Floquet sectors (red and blue panels). (c) Floquet spectrum as function of driving amplitude $\mathcal{E}$. The exchange interaction is extracted from the amplitude dependent singlet-triplet splitting $\epsilon_T-\epsilon_S$.
\label{f:floq}}
\end{figure}

The above two methods for evaluation of nonequilbrium exchange interactions are in principle applicable to arbitrary electric fields. However, they do not give much analytical insight in how strong the exchange interactions can be modified. Considerable insight into this problem can be obtained by considering electric fields adjusted such that the electronic distribution is hardly changed. In this case, it becomes possible to generalize the perturbative result $J_\mathrm{ex}=2t_0^2/U$, as was first demonstrated in \cite{mentink2015} for time-periodic fields with frequencies non-resonant to direct electronic excitations. Considering a simple two-site Hubbard model, the equilibrium exchange interaction can be directly inferred form the singlet-triplet splitting: $J_\mathrm{ex}=(E_T-E_S)/2$, as illustrated in Fig.~\ref{f:floq}a for the $S_z=0$ sector of the two-site Hubbard model at half-filling. Under time-periodic electric fields, one can apply Floquet's theorem \cite{Floquet1883,GrifoniHanggi1998}. Solutions of the time-dependent Schr\"odinger equation are given in the form  $| \psi (t) \rangle = e^{- \text{i}\epsilon_\alpha t } | \psi_\alpha (t) \rangle$ where $| \psi_\alpha (t+T) \rangle = | \psi_\alpha (t) \rangle$ is time-periodic with a period $T=2\pi/\omega$, and  $\epsilon_\alpha$ is a quasi-energy defined up to multiples of $\omega$. A small part of the extended Floquet spectrum of the two-site system is illustrated in Fig.~\ref{f:floq}b. In addition to the virtual hoppings that determine the equilibrium $J_\mathrm{ex}$ (blue arrows), there is virtual absorption and emission of photons to different Floquet sectors (red arrows). The mixing between these Floquet sectors results in a renormalization of the quasi-energy levels. The effective exchange interaction that emerges in the presence of the time-periodic field can then be extracted from the amplitude dependent singlet-triplet splitting $\epsilon_T-\epsilon_S$ (Fig.~\ref{f:floq}c). For the single-band Hubbard model, the amplitude $E_0$ of the external field enters in calculations as the dimensionless driving strength $\mathcal{E}= eaE_0/\hbar \omega$, where $e$ and $a$ are unit charge and lattice spacing, respectively. For weak driving $\mathcal{E}\ll 1$, one obtains $J_\text{ex}=2t_{0}^2/U + \Delta J_\text{ex}$ \cite{mentink2015} with 
\begin{equation}\label{pertb} 
\Delta J_\text{ex} = \frac{\mathcal{E}^2t_{0}^2}{2}\Big(
\frac{1}{U+\hbar\omega}
+
\frac{1}{U-\hbar\omega}
-\frac{2}{U}
\Big).
\end{equation}
Hence, depending on wether the frequency $\hbar\omega$ is below (above) the gap energy $U$, the net effect is an enhancement (reduction) of $J_\text{ex}$. For stronger driving strength we have $J_\text{ex}(\mathcal{E},\omega)=\sum_{m=-\infty}^\infty \frac{2t_0^2J_{|m|}(\mathcal{E})^2}{U+m\hbar\omega}$ Interestingly, when $\mathcal{E}$ is of order one the terms with $U-m\hbar\omega$ can become strongly enhanced, leading to a sign reversal of $J_\text{ex}$. While originally derived on the basis of the two-site model, these results were soon confirmed based on more rigorous time-dependent canonical transformation techniques \cite{itin2015,bukov2016,kitamura2016}, illustrating that the essential physics is captured already within this simple model.

In addition, very similar results can be obtained for a three-site model, which takes into account explicitly that the hopping proceeds via an intermediate uncorrelated orbital and can be regarded as minimal model for superexchange \cite{mikhaylovskiy2015}
\begin{equation}\label{superexchange} 
\Delta J_{ex}
=
\frac{\mathcal{E}^2 t_{0}^4 }{2}\Big\{
\sum_{\pm}
\Big[
\frac{1}{U_1\pm\hbar\omega}
+
\frac{1}{U_1}
\Big]^2
\frac{1}{U\pm\hbar\omega}
-
\frac{4
}{U_1^2U}
-
\frac{4}{U_1^3}
\Big\}.\end{equation}
Here $U_1=U+\Delta$, with $\Delta$ the energy splitting of the uncorrelated orbital with respect to the magnetic sites. The perturbative expressions Eqs.~(\ref{pertb})-(\ref{superexchange}) bear a close analogy with the derivation of the two-magnon Raman scattering Hamiltonian \cite{elliot1963,fleury1968} from the Hubbard model \cite{shastry1990,lorenzana1995prb,devereaux2007}, see also Sec.~\ref{s:2M}. The validity and usefulness of the Floquet picture are discussed in Sec.~\ref{s:driving} by direct comparison with the evaluation of the more general formulas introduced above. 
	
\section{Ultrafast control of $J_\text{ex}$}\label{s:controljex}

In principle, laser excitation can affect $J_\text{ex}$ directly by modulating the electronic structure (electron hopping, Coulomb repulsion) and by creating a nonequilibrium distribution of photoexcited carriers (photodoping). In the latter case, this allows us to address the fundamental question how much time it takes before a static $J_\text{ex}$ emerges from the full electronic dynamics and how much this $J_\text{ex}$ is modified as compared to the equilibrium situation. Alternatively, the goal of modulating the electronic structure, is to achieve a control of $J_\text{ex}$ which is reversible on ultrafast timescales, \textit{i.e.} to give a controlled perturbation to $J_\text{ex}$ during the application of the laser pulse, but leave the electronic state unexcited after the pulse is switched off. Both types of controlling $J_\text{ex}$ are discussed below.

\begin{figure}[h]
\includegraphics[width=\columnwidth]{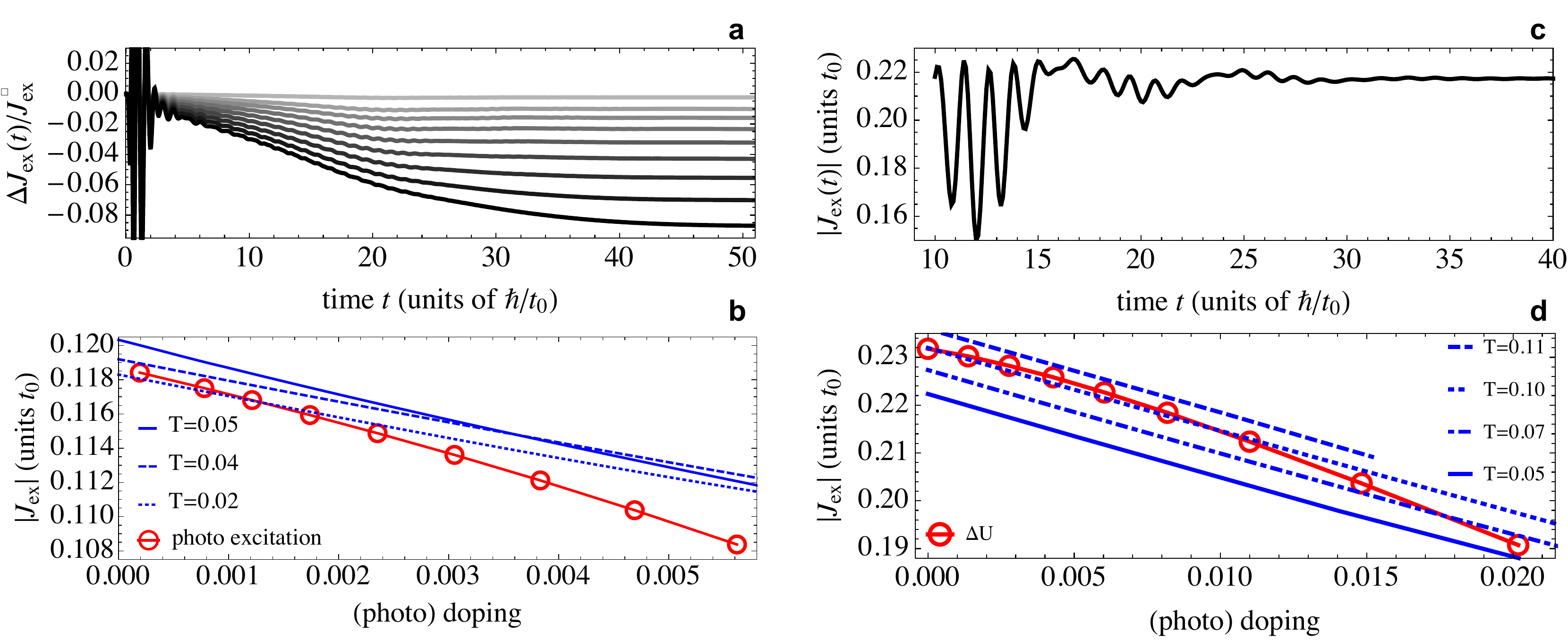}
\caption{
{\bf Ultrafast control of $J_\text{ex}$ by photo doping in a Mott insulator}. (a) time evolution of $\Delta J_\text{ex}(t)$ caused by excitation with an electric field pulse with strengths increasing from $|eaE_0/t_0|=1\ldots 5.5$ (hypercubic lattice in canted geometry, $U/t_0=8$). (b) comparison of the quasistationary nonequilibrium exchange interaction (red open circles) obtained from (a), with the equilibrium exchange interaction in the chemically doped system (blue lines). (c) time-dependent $J_\text{ex}$ evaluated from the general formulas for a quench $U/t_0=4\rightarrow8$. (d) comparison as in (b) obtained using the general formulas for different interaction quenches $\Delta U$. Both (c) and (d) are obtained on the Bethe lattice, with $U/t_0=8$ for the equilibrium results at chemical doping. Photo doping is here defined as the induced change $\Delta n = d + h - d_0 - h_0 = 2 (d-d_0)$ of the doublon and hole densities $d$ and $h$ with respect to their equilibrium values $d_0$ and $h_0$. Reproduced with permission from \cite{mentink2014}, Copyright (2014) by The American Physical Society. 
\label{f:controljex}}
\end{figure}

\subsection{Photo-excitation}\label{photoexcitation}
First we discuss the case of creating a nonequilbrium electronic distribution (photodoping), which was simulated for the AFM Mott insulator using nonequilibrium DMFT \cite{mentink2014}. Results were obtained by evaluating the nonequilibrium $J_\text{ex}$ in the canted geometry (Sec.~\ref{s:jexcanted}) as well as by evaluating the general nonequilibrium exchange formulas (Sec.~\ref{s:jexgeneral}). For the canted geometry, the DMFT equations were solved on the hypercubic lattice and the dynamics was evaluated in response to a single-cycle pulse with center frequency $\hbar\omega=U$. The evolution of $\Delta J_\text{ex}(t)$ is shown in Fig.~\ref{f:controljex}a, which demonstrates that a stationary modified $J_\text{ex}$ emerges already within a few tens of electron hopping times. A very similar time scale is found based on the general exchange formula (Eq.~(\ref{jex:bare})), as shown in Fig.~\ref{f:controljex}c (solid line). In this case, the DMFT equations were solved in a collinear setup on the Bethe lattice and a sudden change (quench) of the Coulomb interaction $U$ was used to create a non-equilibrium electron distribution. Furthermore, it is found that the efficiency of modifying $J_\text{ex}$ is determined by the number of photo-excited carriers, comparable to that of chemical doping. This is shown in the bottom panels, Fig.~\ref{f:controljex}b,d, by plotting the extracted exchange interaction in the quasistationary state as a function of the photodoping, together with equilibrium calculations with chemical doping with the same total number of carriers.

\subsection{Modulating the electronic structure}\label{s:driving}
Second we discuss the control of $J_\text{ex}$ by modulating the electronic structure. In principle, the Floquet picture gives the effective exchange interaction under strictly periodic fields. However, in the long time limit isolated many-body systems can become infinitely excited and an effective low-energy description of the system is of limited use \cite{dalessio2014,lazarides2014}. Moreover, for short pulses with only a few number of cycles, it is \textit{a priori} not clear how accurate the Floquet picture is. Below we review simulation results, obtained with both exact diagonalization and nonequilibrium DMFT, which nevertheless confirm the Floquet picture, at least for the short-time dynamics on the time-scale defined by $\hbar/J_\text{ex}$.

A direct confirmation of Eq.~(\ref{superexchange}) was obtained by evaluating Eq.~(\ref{jex:bare}) on a three-site cluster driven by a time-periodic field with an amplitude that is slowly ramped on, within about 10 cycles, while the electron dynamics is solved using exact diagonalization \cite{mikhaylovskiy2015}. The effective exchange interaction was extracted by averaging over the period of the field:
\begin{equation}
J^\mathcal{E}_{ij}(t) = \frac{1}{T}\int_{t}^{t+T}\!\!\!\!\!\!ds\,J_{ij}(s),
\end{equation}
and found to be quasi-stationary for times sufficiently long after the ramp. Excellent quantitative agreement was obtained for the ratio $\Delta J/J\propto \mathcal{E}^2$ in the regime $\mathcal{E}\ll1$. This is the regime that is most relevant to experiments on condensed matter systems and it was estimated that for model parameters typical for experiments on iron oxides a relative change $\Delta J/J\sim1\%$ is achieved.

\begin{figure}[h]
\includegraphics[width=\columnwidth]{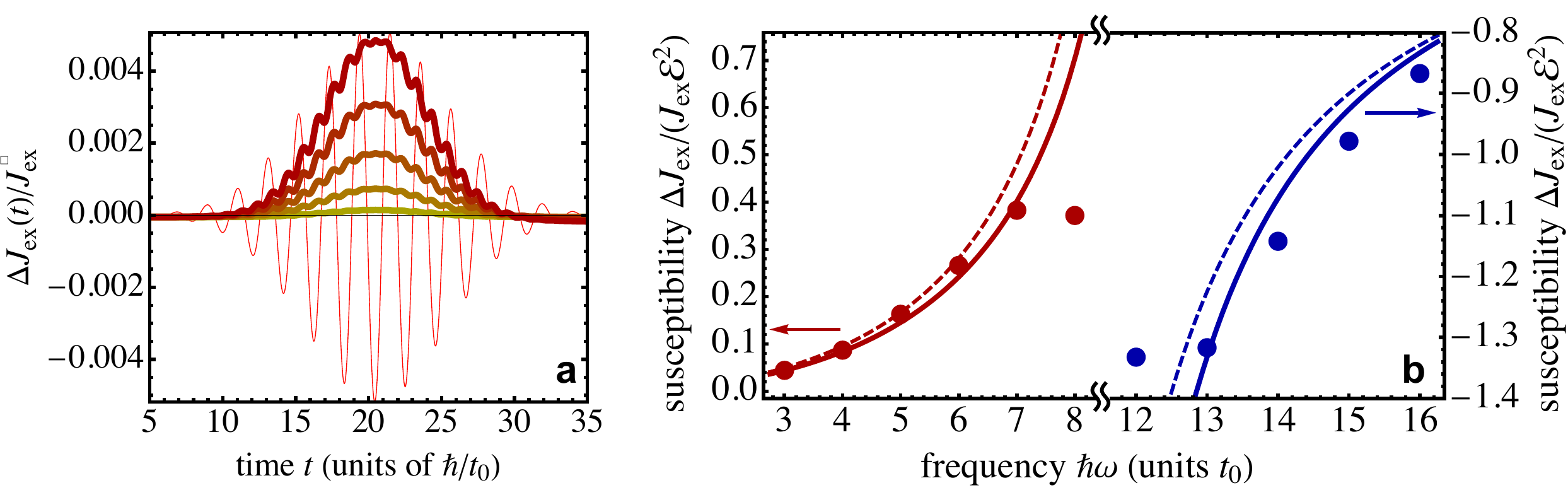}
\caption{
{\bf Control of $J_\text{ex}$ by periodic driving}. (a) Time-dependent change of the period-averaged exchange interaction ($\Delta J_\text{ex}$, thick lines) during the action of an oscillatory electric field pulse (thin lines), with driving frequency $\hbar\omega=3$ below the Mott gap. Different colours correspond to results obtained with different amplitude $E_0$ of the electric field, increasing from light to dark. Numerical results were obtained using Dynamical Mean Field Theory (DMFT) for the hyper-cubic lattice at $U=10$ and initial temperature $T=0.025$. (b) The driving susceptibility $\Delta J_\text{ex}/(J_\text{ex}\mathcal{E}^2)$ for $\mathcal{E}\to0$ for frequencies above (blue, right vertical axis) and below gap (red, left vertical axis), obtained from DMFT for the hyper-cubic lattice (disks), from the numerical Floquet spectrum of a two-site Hubbard cluster (solid lines), and from the perturbative result Eq.~(\ref{pertb}) (dashed lines). Reproduced with permission from \cite{mentink2015}.
\label{f:jexdriving}}
\end{figure}

In a similar way, the period-averaged $J_\text{ex}$ was compared in the canted DMFT setup \cite{mentink2015}. In this case, a Gaussian envelope function was used containing about 15 cycles. For driving frequencies sufficiently far away from resonance, it is found that $\Delta J_\text{ex}\approx 0$ after the pulse (see Fig.~\ref{f:jexdriving}a, where $\hbar\omega=3$ and $U=10$), demonstrating that the electronic state of the system remains unexcited after the pulse has left and hence the exchange interaction can be controlled reversibly on ultrafast time scales. Furthermore, in accordance with the Floquet prediction, in the perturbative regime an enhancement (reduction) of $J_\text{ex}$ is obtained for driving below (above) gap. This is shown in Fig.~\ref{f:jexdriving}b, by a quantitative comparison with the Floquet theory for the "driving susceptibility" $\Delta J_\text{ex}/\left(J_\text{ex}\mathcal{E}^2\right)$. The solid symbols are obtained from running several DMFT calculations at increasing field strength (different colors in Fig.~\ref{f:jexdriving}a), while the dashed and solid lines are based on the perturbative result (Eq.~(\ref{pertb}) and the full Floquet spectrum (non-perturbative in $t_0/U$) by evaluating the derivative $dJ_\text{ex}/d\mathcal{E}^2$ at $\mathcal{E}\rightarrow0$, respectively. Away from the band edge, where the system is photo-excited (Sec.~\ref{photoexcitation}), the frequency dependence matches very well, being even in quantitative agreement for the lowest frequencies below gap. Hence, the Floquet theory forms a useful guide for understanding the ultrafast and reversible control of $J_\text{ex}$ in condensed matter systems by photo-assisted hopping.

\section{Manipulation of magnetism}
Modelling how magnetic order can be manipulated by short time-dependent perturbations of exchange interactions can be addressed on several levels, four of which are discussed here. First, we focus on macrospin theory, which is suitable to describe homogenous spin precession and was used to establish the link between the experimentally observed AFM resonances and the sub-picosecond control of exchange interactions \cite{mikhaylovskiy2015}. Second, atomistic spin dynamics simulations are used to investigate the response of the spin temperature to sudden changes in $J_\text{ex}$ \cite{hellsvik2016}. Third, harmonic magnon theory is used to model impulsively stimulated two-magnon Raman scattering due to perturbations of $J_\text{ex}$, leading to longitudinal macrospin dynamics \cite{bossini2016,zhao2004,zhao2006}. Finally, the quantum spin dynamics of a one-dimensional chain is studied, leading to effective time reversal under influence of a change of sign of $J_\text{ex}$ by periodically modulating the electronic structure \cite{mentink2015}. 

\subsection{Excitation of spin precession}\label{s:afmr}
Modeling of macroscopic spin precession can be conveniently done by solving the multi-scale problem on the basis of the macrospin approximation. In this case, each of the magnetic sublattices is treated as an effective macrospin with dynamics governed by the Landau-Lifshitz equation. As introduced in Sec.~\ref{s:jexcanted}, we can also use the Landau-Lifshitz equation to infer an effective $J_\text{ex}(t)$ from an observed spin precession. Here, we focus how this approach was used to provide experimental evidence for an ultrafast control of exchange interactions.

Instead of a canting induced by an external magnetic field, the experiments were performed on iron oxides that are intrinsically canted due to an additional antisymmetric, so-called Dzyaloshinskii-Moriya interaction and described by $H_\text{DM}=\mathbf{D}\cdot\sum_{ij}\left(\mathbf{S}_i\times\mathbf{S}_j\right)$. General symmetry arguments can be used to prove that perturbations to both $J_\text{ex}$ and $\mathbf{D}$ contain an intensity dependent contribution ($\propto |\mathbf{E}|^2$). For example, by using Eq.~(\ref{pertb}) and assuming a simple cubic lattice, this also follows directly from the microscopic model. In the macrospin approximation, the torque on $\langle\mathbf{S}_1\rangle$ due to $\langle\mathbf{S}_2\rangle$ is given by $\mathbf{T}_1= - \langle\mathbf{S}_1\rangle\times\Delta\mathbf{B}_\text{ex}$, where
\begin{eqnarray}
\Delta\mathbf{B}_\text{ex} &= \sum_j \Delta J_{1j}\langle\mathbf{S}_2\rangle = \left(2\alpha E_x^2 + 2\alpha E_y^2 + 2\alpha E_z^2\right)\langle\mathbf{S}_2\rangle = 2\alpha |\mathbf{E}|^2\langle\mathbf{S}_2\rangle,
\end{eqnarray}
where the summation is over the six nearest neighbor bonds. Defining $\mathbf{E}=E_0\mathbf{e}$, $\mathbf{e}$ the unit vector of polarization, we recover $\alpha|\mathbf{E}|^2=\Delta J_\text{ex}$, with $\Delta J_\text{ex}$ defined by Eq.~(\ref{pertb}). Hence, although only those exchange bonds are perturbed that have a projection along the electric field, it follows that the torque, which is a sum over all bonds, is independent on the polarization of light. This isotropy is in strong contrast with previously reported mechanisms for the optical excitation of spin resonances, such as the inverse Faraday \cite{kimel2005} and inverse Cotton-Motton effect \cite{kalashnikova2007}, which depend on the helicity of light and on the orientation of the polarization with respect to the magnetization, respectively. Hence, the key experimental signature of optical perturbation of exchange interactions, is an isotropic, polarization independent excitation of the spin resonance. This is exactly what has been observed in the experiments on femtosecond laser excitation of iron oxides, where the subsequent spin dynamics was detected using THz emission spectroscopy \cite{mikhaylovskiy2015}. A quantitative analysis supports the sub-picosecond time scale at which $\Delta J_\text{ex}$ can occur, consistent with the theoretical results discussed in Sec.~\ref{s:controljex}. Moreover, in the same and similar materials not only the control of $J_\text{ex}$ between transition metal ions was found, but also the first indications were reported for the control of $J_\text{ex}$ between the rare-earth ions \cite{mikhaylovskiy2015prb,mikhaylovskiy2017}. We emphasize, however, that the experiment \cite{mikhaylovskiy2015} is sensitive to an ultrafast perturbation of the \emph{ratio} $|\mathbf{D}|/J_\text{ex}$ and hence does not provide a direct proof of the modification of $J_\text{ex}$ alone. 

\subsection{Cooling by perturbation of exchange}
\begin{figure}[h]
\center{\includegraphics[width=0.7\columnwidth]{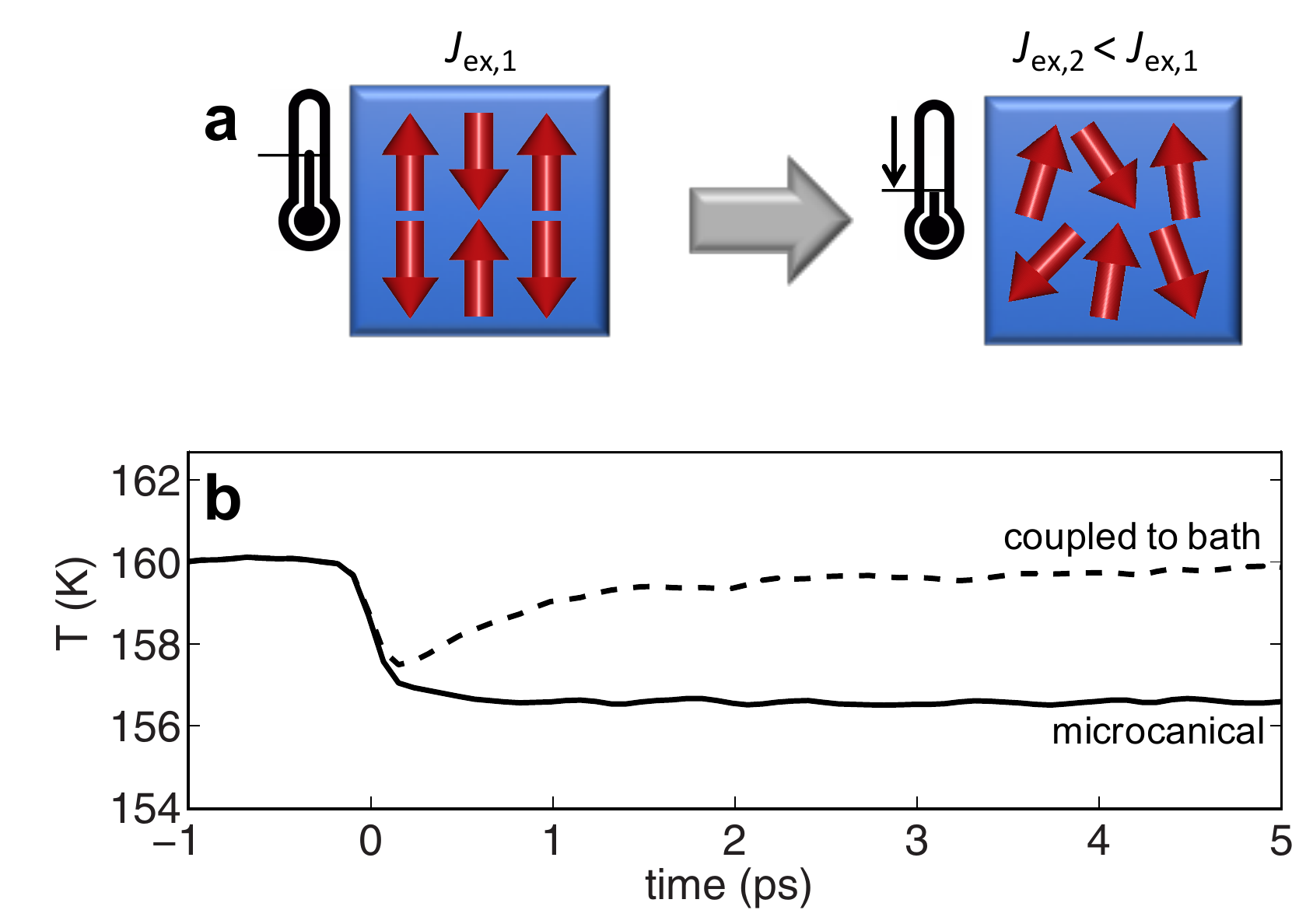}}
\caption{
{{\bf Rapid cooling by perturbations of exchange interactions.} (a) Illustration of cooling by reduction of $J_\text{ex}$. (b) results of atomistic spin dynamics simulations where the response of the spin temperature is measured after a step-like change of exchange interactions, as modelled by the introduction of a fraction of $x=0.02$ vacancies at $t=0$~ps. Solid lines show the microcanonical evolution, dashed lines include coupling to the bath, with a dimensionless coupling strength $\alpha=0.01$. Reproduced with permission from \cite{hellsvik2016}. Copyright (2016) by The American Physical Society.} 
\label{f:cooling}}
\end{figure}

Perturbation of exchange interactions can also influence the relaxation and internal equilibration of the spin degrees of freedom. To simulate such effects, the dynamics of an ensemble of spins has be be solved. In the regime where a classical description is valid, this can be done conveniently on the basis of atomistic spin dynamics (ASD) \cite{reviewyork,bookuppsala}. Within ASD, the dynamics of each atomic spin in the system evolves according to the stochastic Landau-Lifshitz equation
\begin{equation}
\hbar\frac{d\mathbf{S}_i}{dt}=-\mathbf{S}_i\times\big(\mathbf{B}_i+\mathbf{B}^\text{fl}_i(t)\big) - \frac{\alpha}{S_i}\mathbf{S}_i\times\big[\mathbf{S}_i\times\big(\mathbf{B}_i+\mathbf{B}^\text{fl}_i(t)\big)\big]
\end{equation}
which describe the motion of the classical spins $\mathbf{S}_i$ in an effective magnetic field $\mathbf{B}_i$, calculated from $\mathbf{B}_i =-\partial H_\text{spin}/\partial\mathbf{S}_i$. $\mathbf{B}^\text{fl}_i(t)$ is a stochastic magnetic field with a Gaussian distribution. By the fluctuation-dissipation theorem, the magnitude of $\mathbf{B}^\text{fl}_i(t)$ is related to the dimensionless damping parameter $\alpha$. ASD simulations in response to perturbations of exchange interactions were carried out for CuO \cite{hellsvik2016} based on equilibrium exchange parameters computed from first principles, which involves both bilinear Heisenberg exchange and biquadratic exchange interactions \cite{hellsvik2014}. Fig.~\ref{f:cooling} shows the evolution of the temperature of the system as evaluated by fitting a Boltzmann distribution to the energy distribution of the spins, after introducing vacancies (sites without spin) with a concentration $x=0.02$. Interestingly, starting in the low-temperature collinear antiferromagnetic phase, the system evolves within the AFM phase by rapid internal equilibration which reduces the temperature. This can be explained as an ultrafast magneto-caloric effect, where the closed system follows a constant entropy curve in response to lowering the strength of the exchange fields. This interpretation was confirmed by comparison with the equilibrium entropy per spin at different concentrations $x$, showing close to quantitative agreement. When the system is coupled to a heat bath, still a transient drop is observed, after which the system relaxes to the temperature of the heat bath (dashed line in Fig.~\ref{f:cooling}). For small $x$, qualitatively similar results were obtained by a step-like reduction of exchange parameters. Although several aspects of existing experimental work \cite{johnson2012} on laser-induced dynamics in CuO are not captured by the simulations shown here (see \cite{hellsvik2016} for a detailed discussion), the results do suggest new opportunities to achieve ultrafast laser-induced cooling of magnets by optical perturbations of exchange, strongly contrasting the laser-induced heating commonly observed in metallic magnets \cite{kirilyuk2010}.

\subsection{Excitation of coherent longitudinal spin dynamics}\label{s:2M}
In collinear antiferromagnets, perturbations of exchange interactions do not give rise to excitation of the homogenous antiferromangetic resonance (AFMR) mode, since the effective field generated by the perturbation of $J_\text{ex}$ is collinear with the sublattice magnetization and therefore the torque vanishes: $\mathbf{T}_1=-\langle\mathbf{S}_1\rangle\times2\Delta J_\text{ex}\langle\mathbf{S}_2\rangle=0$. Nevertheless, it was recently reported that coherent $\textit{longitudinal}$ oscillations of the AFM order parameter can be triggered by ultrashort perturbations of exchange interactions \cite{bossini2016}. Modeling of such longitudinal dynamics goes beyond the macrospin approximation employed in Sec.~\ref{s:afmr}, which conserves the length of the sublattice magnetization $|\langle \mathbf{S}_i\rangle|$. Extending on previous results \cite{zhao2004,zhao2006}, it was found that the longitudinal dynamics emerges naturally within harmonic magnon theory \cite{bossini2016} due to the impulsive excitation of pairs of magnons with opposite $\mathbf{k}$ by perturbations of $J_\text{ex}$. Here we discuss these theoretical results by showing that they can be understood from a simple quantum-mechanical two-level system. 

Following \cite{zhao2004,bossini2016}, we start from an unperturbed Heisenberg Hamiltonian $H_0=J_\text{ex}\sum_{i\boldsymbol{\delta}}\mathbf{S}_i\mathbf{S}_{i+\boldsymbol{\delta}}$. Perturbations to $J_\text{ex}$ depend on the orientation of the electric field and for the simple cubic lattice it follows from Eq.~(\ref{pertb}) that we can write $\delta H = \Delta J_\text{ex}\sum_{i\boldsymbol{\delta}}(\mathbf{e}\cdot\boldsymbol{\delta})^2 \mathbf{S}_i\mathbf{S}_{i+\boldsymbol{\delta}}$. Here $\boldsymbol\delta$ is the unit vector connecting two adjacent sites on different sublattices and $\mathbf{e}$ is the unit vector of the polarization of light. More generally, one can use here also the phenomenological second order Raman tensor and the present description is analogous to the one used for impulsive Raman scattering of phonons \cite{stevens2002}. Magnons are described in the usual way by introducing Holstein-Primakov bosons $a_\mathbf{k},b_\mathbf{k}$ for spins in different sublattices \cite{kittel1963,fazekas1999}. Keeping only the terms bilinear in the magnon operators we have
\begin{eqnarray}
H_0'&= zJ_\text{ex}S\sum_\mathbf{k}\left[\gamma_\mathbf{k}\big(a_\mathbf{k}b_{-\mathbf{k}}+a^\dagger_\mathbf{k}b^\dagger_{-\mathbf{k}}\big) + \big(a_\mathbf{k}^\dagger a_{\mathbf{k}}+b^\dagger_{-\mathbf{k}}b_{-\mathbf{k}}\big)\right],\\
\delta H'&= z\Delta J_\text{ex}S\sum_\mathbf{k}\left[\xi_\mathbf{k}\big(a_\mathbf{k}b_{-\mathbf{k}}+a^\dagger_\mathbf{k}b^\dagger_{-\mathbf{k}}\big) + \big(a_\mathbf{k}^\dagger a_{\mathbf{k}}+b^\dagger_{-\mathbf{k}}b_{-\mathbf{k}}\big)\right].
\end{eqnarray}
Here $\gamma_\mathbf{k}=\frac{1}{z}\sum_{\boldsymbol{\delta}}\exp({i\mathbf{k}\cdot\boldsymbol{\delta}})$ and $\xi_\mathbf{k}=\frac{1}{z}\sum_{\boldsymbol{\delta}}(\mathbf{e}\cdot\boldsymbol{\delta})^2\exp({i\mathbf{k}\cdot\boldsymbol{\delta}})$. Due to the exchange interaction between spins of different sublattices, $H_0$ is not diagonal in the magnon operators $a_\mathbf{k},b_\mathbf{k}$. To diagonalize $H_0$, composite magnons are introduced by a Bogoliubov transform
\begin{eqnarray}
a_\mathbf{k}&=u_\mathbf{k}\alpha_\mathbf{k} + v_\mathbf{k}\beta^\dagger_{-\mathbf{k}},\\
b_\mathbf{k}&=u_\mathbf{k}\beta_\mathbf{k} + v_\mathbf{k}\alpha^\dagger_{-\mathbf{k}}.
\end{eqnarray}
where the coefficients $u_\mathbf{k},v_\mathbf{k}$ are chosen such that the off-diagonal terms vanish, yielding $H_0'=\sum_\mathbf{k}\hbar\omega_\mathbf{k}\left(\alpha_\mathbf{k}^\dagger\alpha_\mathbf{k} + \beta_\mathbf{k}^\dagger\beta_\mathbf{k}\right)$, where $\hbar\omega_\mathbf{k}=zJ_\text{ex}S\sqrt{1-\gamma_\mathbf{k}^2}$. However, the same transformation does not diagonalize $\delta H'$:
\begin{eqnarray}\label{e:dh}
\delta H'&= \sum_\mathbf{k}\delta\hbar\omega_\mathbf{k}\left(\alpha_\mathbf{k}^\dagger\alpha_\mathbf{k} + \beta_\mathbf{k}^\dagger\beta_\mathbf{k}\right) + V_\mathbf{k}\left(\alpha_\mathbf{k}\beta_{-\mathbf{k}} + \alpha^\dagger_\mathbf{k}\beta^\dagger_{-\mathbf{k}}\right),
\end{eqnarray}
where $\delta\hbar\omega_\mathbf{k}=z\Delta J_\text{ex}S\left(1-\xi_\mathbf{k}\gamma_\mathbf{k}\right)/\sqrt{1-\gamma_\mathbf{k}^2}$, $V_\mathbf{k}=z\Delta J_\text{ex}S\left(\xi_\mathbf{k}-\gamma_\mathbf{k}\right)/\sqrt{1-\gamma_\mathbf{k}^2}$. Since in general $\xi_\mathbf{k}\neq\gamma_\mathbf{k}$, pairs of magnons with opposite $\mathbf{k}$ are excited due to the second term in Eq.~(\ref{e:dh}). From the structure of this term, we see that the response to time-dependent perturbations $\Delta J_\text{ex}(t)$ maps onto the solution of a collection of independent two-level systems for each pair $\mathbf{k},-\mathbf{k}$. In particular, starting from the ground state $|g_\mathbf{k}\rangle=|0\rangle|0\rangle$, $\delta H'$ only induces couplings to excited states $|e_\mathbf{k}\rangle=\alpha^\dagger_\mathbf{k}\beta^\dagger_{-\mathbf{k}}|0\rangle|0\rangle$ and we have a two-level system specified by 
\begin{eqnarray}
\langle g_\mathbf{k}|H_0'+\delta H'| g_\mathbf{k}\rangle &= 0\\
\langle e_\mathbf{k}|H_0'+\delta H'| e_\mathbf{k}\rangle &= 2\big[\hbar\omega_\mathbf{k}+\delta\hbar\omega_\mathbf{k}(t) \big]\\
\langle g_\mathbf{k}|H_0'+\delta H'| e_\mathbf{k}\rangle &= \langle e_\mathbf{k}|H_0'+\delta H'|g_\mathbf{k} \rangle^* = V_\mathbf{k}(t).
\end{eqnarray}
The solution can be obtained by writing $|\psi_\mathbf{k}(t)\rangle = c_\mathbf{k}(t)|g_\mathbf{k}\rangle + d_\mathbf{k}(t)|e_\mathbf{k}\rangle$. For impulsive excitation we write $V_\mathbf{k}(t)=\tau\hbar\overline{V_\mathbf{k}}\delta(t)$ and obtain analytical expressions valid for $t\gg\tau$:
\begin{eqnarray}
c_\mathbf{k}(t) &= c_\mathbf{k}(0)\cos (\tau\overline{V_\mathbf{k}}),\quad d_\mathbf{k}(t) &= -ic_\mathbf{k}(0)\sin (\tau\overline{V_\mathbf{k}}) e^{i2\omega_\mathbf{k}t},
\end{eqnarray}
where we for simplicity neglected a small phase accumulated due to $\delta\hbar\omega_\mathbf{k}(t)$. Longitudinal dynamics follows by evaluating 
\begin{eqnarray}
\langle L_z(t)\rangle&=\sum_k\langle\psi_\mathbf{k}(t)|\hat{L}_z|\psi_\mathbf{k}(t)\rangle = NS - \frac{2}{N}\sum_\mathbf{k}\langle\psi_\mathbf{k}(t)|a_\mathbf{k}^\dagger a_\mathbf{k}+b_\mathbf{k}^\dagger b_\mathbf{k}|\psi_\mathbf{k}(t)\rangle,
\end{eqnarray}
where $N$ is the total number of spins and $\hat{L}_z=\sum_{i,j}\left(S_{iz}-S_{j z}\right)$ is the $z-$component of the antiferromagnetic vector $\mathbf{L}$. Substituting the Bogoliubov transformation and keeping only terms linear in $\overline{V}_\mathbf{k}$ ($\Delta J_\text{ex}\ll J_\text{ex}$), we find that the dynamics of $\langle L_z(t)\rangle$ is determined by the terms $\langle\psi_\mathbf{k}(t)|\alpha_\mathbf{k}\beta_{-\mathbf{k}}+\alpha_\mathbf{k}^\dagger \beta^\dagger_{-\mathbf{k}}|\psi_\mathbf{k}(t) \rangle=c_\mathbf{k}^*(0)d_\mathbf{k}(t)+c_\mathbf{k}(0)d_\mathbf{k}^*(t)=-2c^2_\mathbf{k}(0)\overline{V_\mathbf{k}}\sin(2\omega_\mathbf{k}t)$, finally giving
\begin{eqnarray}
\langle L_z(t)\rangle&=\text{const}-2\sum_\mathbf{k}\frac{\tau\overline{V_\mathbf{k}}c_\mathbf{k}^2(0)\gamma_\mathbf{k}}{\sqrt{1-\gamma_\mathbf{k}^2}}\sin(2\omega_\mathbf{k}t).
\end{eqnarray}

\begin{figure}[h]
\raggedleft{\includegraphics[width=0.8\columnwidth]{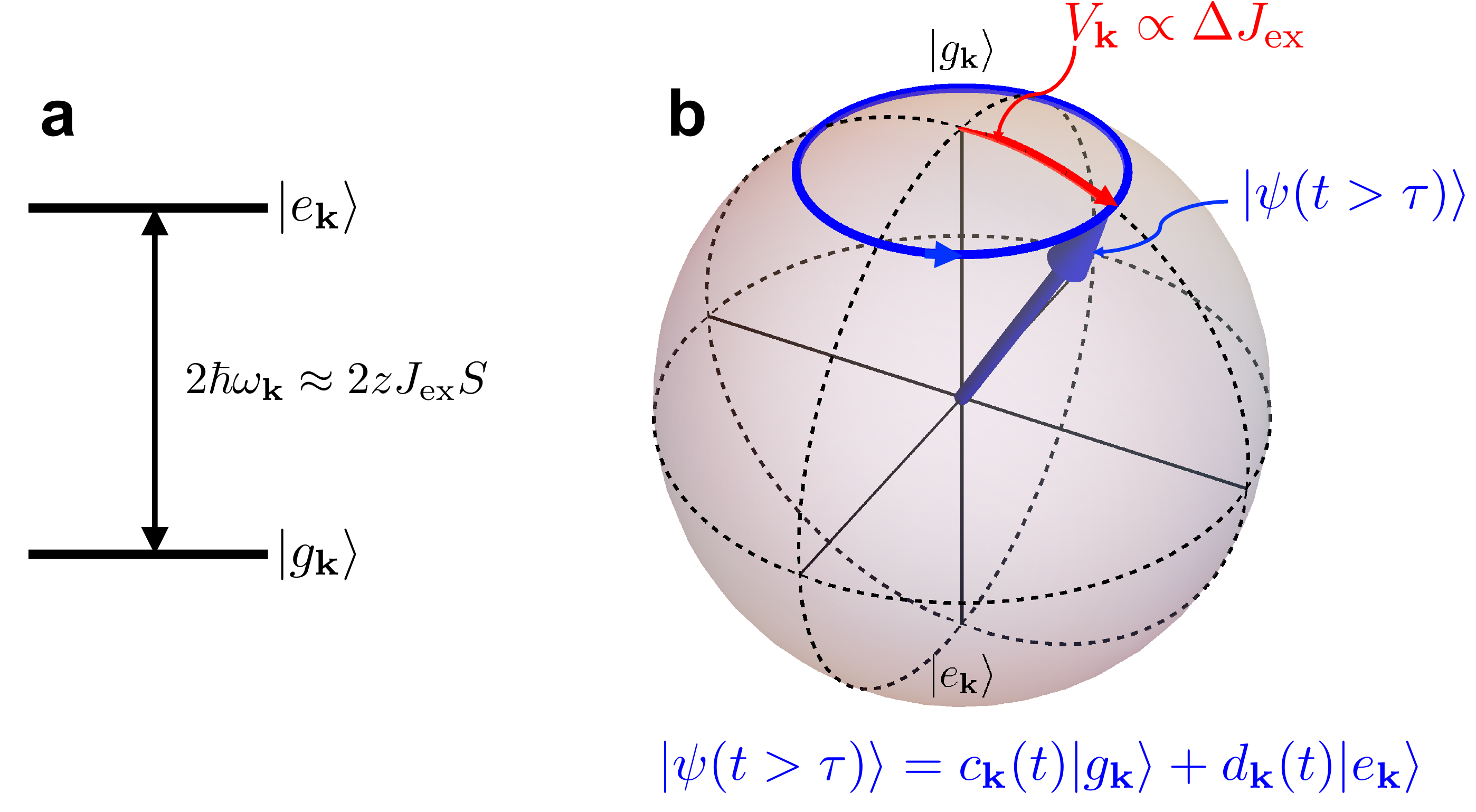}}
\caption{
(a) Two-level system with approximate splitting $2\hbar\omega_\mathbf{k}\propto2J_\text{ex}$. (b) Bloch sphere representation of the two-level system with north and south pole representing the ground and two-magnon excited state, respectively. The red line indicates the effect of the  perturbation $V_\mathbf{k}\propto \Delta J_\text{ex}$ which brings the system in a super position of the ground sate and excited state (blue arrow), after which the wave function follows the blue trajectory, oscillating at twice the single magnon frequency. Note that for the two-magnon case, unlike for homogenous spin precession, the projections of $|\psi(t)\rangle$ onto the cartesian axes do not correspond to different components of the homogenous magnetization.
\label{f:2M2level}}
\end{figure}

Although qualitatively the dynamics of the longitudinal dynamics described by this equation is very different from homogenous spin precession, the solution of the problem has a very similar mathematical structure. In particular, due to the mapping onto a collection of independent two-level systems of composite magnons, the solution of the time-dependent wave function can be represented on the Bloch sphere, analogous to the case of homogenous spin precession described by a two-level system \cite{feynman1957,gridnev2008}. This is illustrated in Fig.~\ref{f:2M2level}, where the north and south pole represent the ground and two-magnon excited state, respectively. During the application of the optical pulse, perturbations of exchange interactions bring the system into a super position of the ground and excited state, as indicated by the red line. After the pulse is gone, this coherence  $c_\mathbf{k}^*(0)d_\mathbf{k}(t)+c_\mathbf{k}(0)d_\mathbf{k}^*(t)$ remains and the wave function (blue arrow) follows the blue trajectory on the Bloch sphere. For the two-magnon case, the excited state comprises two magnons and hence the oscillation occurs at twice the single magnon frequency.

In principle magnons of all $\mathbf{k}$ contribute to the dynamics of $L_z$. However, close to the Brillouin zone-boundary the magnon density of states peaks due to the vicinity of a van-Hove singularity. Therefore, magnons close to the zone edge ($\gamma_\mathbf{k}^2\ll1$) give the dominant contribution and $L_z$ oscillates at $2\omega_\mathbf{k}= 2zJ_\text{ex}S\sqrt{1-\gamma_\mathbf{k}^2}/\hbar\approx 2zJ_\text{ex}S/\hbar$. This gives coherent spin dynamics in the femtosecond regime \cite{bossini2016}, which is much faster than the frequency of AFM resonance ($\omega\sim S\sqrt{J_\text{ex} K}/\hbar$, where $K\ll J_\text{ex}$ the anisotropy energy). While we limited ourselves here to a simple two-level description of existing theoretical results, it will be very interesting to better understand this regime of longitudinal spin dynamics and in particular the relation with quantum effects such as the magnon squeezing discussed in earlier works \cite{zhao2004,zhao2006}.

\subsection{Effective time reversal}
\begin{figure}[h]
\includegraphics[width=0.8\columnwidth]{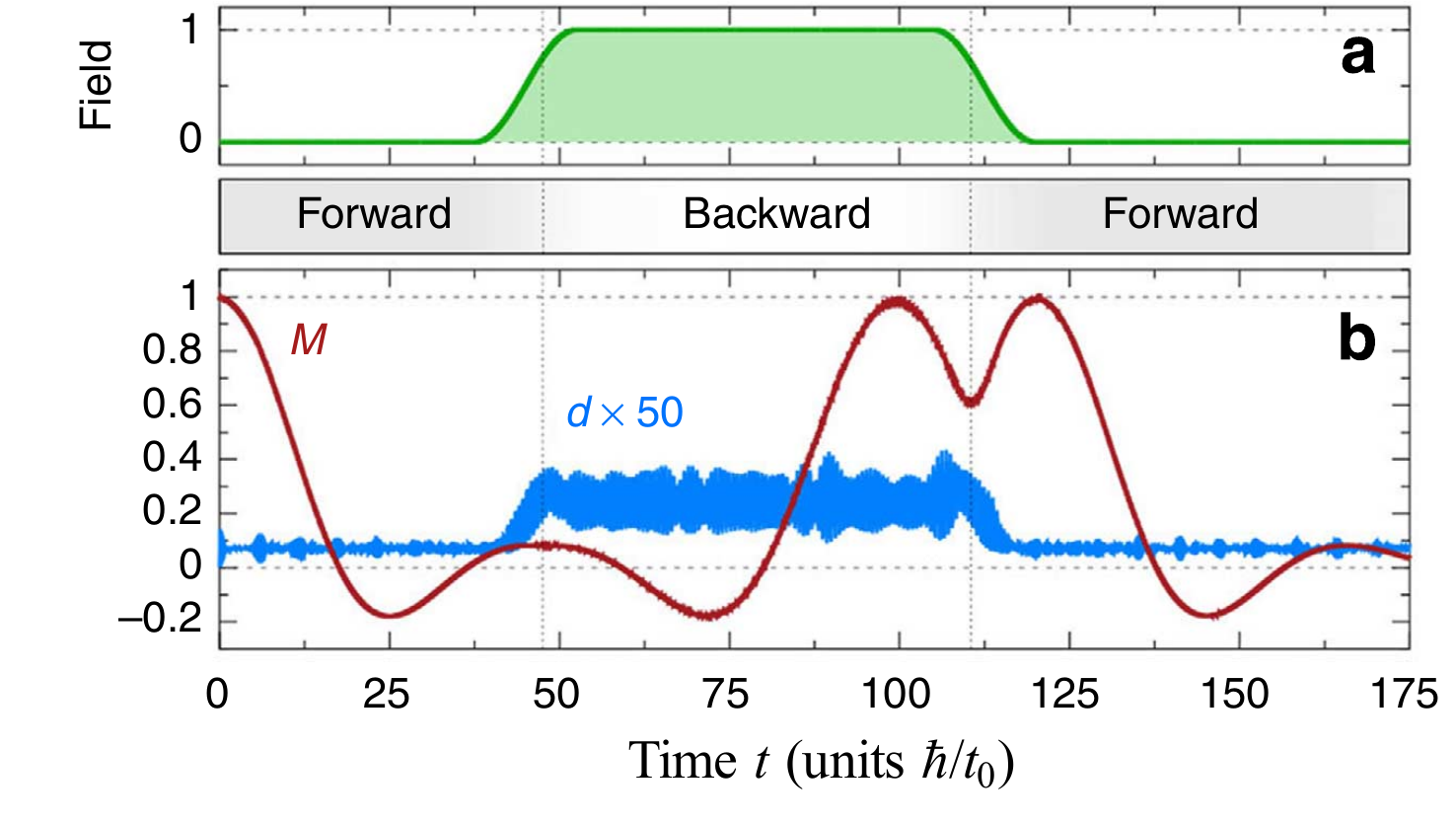}
\caption{
{\bf Effective time reversal of spin dynamics}. (a) Field envelope with cosine-shaped ramps of length $\Delta t=15$ around $t_1= 45$ and $t_2= 112.5$. The bar below the field envelope indicates forward (grey) and backward (white) time evolution when the field is off and on, respectively. (b) Time evolution of staggered magnetization $M$ (red) and total double occupation $d$ (blue), scaled by a factor of 50, of a 10-site Hubbard chain, showing free evolution for times to $t<37.5$ and $t>120$ and evolution under an additional periodic driving at frequency $\omega/U=0.6$ and Floquet amplitude $\mathcal{E}=3.4$ in between. Reproduced with permission from \cite{mentink2015}.
\label{f:timereversal}}
\end{figure}

For sufficiently strong driving strength, the Floquet theory predicts that even the sign of the exchange interaction can be changed. A naive equilibrium analysis would suggest the system evolves into a ferromagnetic state. However, in dynamics this is not possible, since the total spin is conserved under time evolution with the Heisenberg Hamiltonian. Nevertheless, even if the system remains AFM, a change of sign of $J_\text{ex}$ allows for a very non-trivial way to control the spin dynamics, namely, to reverse the time evolution of the undriven system. Here we show the example of the one-dimensional spin chain \cite{mentink2015}. Opposed to the examples discussed before, for small systems the full electronic dynamics of the Hubbard model, including their spin degrees of freedom, is computationally tractable with exact diagonalization. The results of such simulations for a chain of $N=10$ sites are shown in Fig.~\ref{f:timereversal}. To demonstrate the effective time-reversal, the system is prepared in the uncorrelated N\'eel state, which is a highly excited state of the one-dimensional chain and evolution under the unperturbed Hamiltonian will show a rapid decrease of the staggered magnetization $M=\frac{1}{N}\sum_{i=1}^N(-1)^{i+1}\langle\hat{n}_{i\su}-\hat{n}_{i\sd}\rangle$. Subsequently, a time-periodic electric field is ramped on (Fig.~\ref{f:timereversal}a), with Floquet amplitude $\mathcal{E}=3.4$ and frequency $\hbar\omega/U=0.6$, for which the Floquet theory for a two-site model predicts a reversal of the exchange coupling. Under the periodic driving, one indeed observes a near perfect reversal of the dynamics of $M(t)$ in Fig.~\ref{f:timereversal}b, which almost completely recovers the initial value $M(t=0)$ around $t=100$. Subsequently, $M(t)$ is reduced again by further evolution in the reverse direction. This continues until the field is ramped off, after which one observes that the free evolution brings the system again back to the initial state, from which the same rapid decay of $M(t)$ is observed as for the initial free evolution. This result can be well understood from the time-evolution of the pure quantum spin Hamiltonian. Propagation over a time interval $t$ with the unperturbed Hamiltonian is given by the evolution operator $\mathcal{U}_\text{AFM}=\exp(-iH_\text{ex}t/\hbar)$. This evolution is exactly reversed by propagation with $J_\text{ex}'$ of opposite sign over a time interval $t'=|J_\text{ex}/J_\text{ex}'|t$, since for the ferromagnetic (FM) time evolution we have $\mathcal{U}_\text{FM}=\exp(-iH'_\text{ex}t'/\hbar)=\exp(+iH_\text{ex}t/\hbar)=\mathcal{U}_\text{AFM}^{-1}$. Hence, the full dynamics of the Hubbard model very nicely resembles the dynamics expected from the pure spin model with dynamically perturbed $J_\text{ex}$, further confirming the reversibility of the optical control of $J_\text{ex}$. Interestingly, we observe that analogous to the dynamics discussed in Sec.~\ref{s:2M}, the evolution shows coherent purely longitudinal spin dynamics on the time scale determined by $J_\text{ex}$. While the present results may not be directly relevant for condensed matter systems due to the strong field strengths required ($\mathcal{E}\sim1$ corresponds to fields $E_0\sim 1$~V/\AA), they provide novel possibilities to study the reversibility of quantum many body dynamics in cold atom systems.

\section{Conclusion and Outlook}
The aim of this review was to describe the recent theoretical and computation work focused on the description of both the control of exchange interactions under electronic nonequilibrium conditions as well as the response of antiferromagnetic order to such perturbations. Interestingly, besides new ways to excite spin precession, it was found that qualitatively new relaxation and coherent spin dynamics emerges due to ultrafast perturbations of exchange interactions in antiferromagnets. While the simulations on the control of the Heisenberg exchange interaction $J_\text{ex}$ reviewed here are based on the prototype single-band Hubbard model, natural extensions are to study systems with competing interactions, which in general requires studying multi-band systems. Recently, competing exchange interactions have already been studied for effective single-band systems with different underlying lattice geometry and by including higher order terms in the hopping $t_0$ \cite{claassen2016,stepanov2017,kitamura2017}. In addition, also the first studies on multi-orbital systems has been performed \cite{gavrichkov2017,eckstein2017}. Interestingly, beyond the control with laser pulses in the optical regime, \cite{sato2016,meyer2017,eckstein2017} also discuss the control by THz electric field fields and transients, which suggest new opportunities for non-linear spin responses to THz pulses \cite{kampfrath2010,baierl2016a,baierl2016b}, involving both the electric and the magnetic field of the THz pulse. For the manipulation of magnetic order, it will be very interesting to further study the influence of competing exchange interactions for the cooling by perturbations of exchange interactions. This potentially can lead to very efficient cooling strategies, since the exchange fields are much stronger than conventional external magnetic fields, suggesting high potential for magnetic refrigeration applications \cite{bruck2007}. Finally, it will be very interesting to go beyond the limitations of the harmonic magnon theory and the exact diagonalization to characterize and further explore the regime of coherent longitudinal spin dynamics, which keeps great promise for finding new physical phenomena in the short-time dynamics of macroscopic magnetic order out of equilibrium.

\section{Acknowledgement}
The results reviewed in this article have only been possible through inspiring supervision and co-work by M. Eckstein, dedicated co-work with K. Balzer and stimulating collaborations with A. Secchi, M.I. Katsnelson, R.V. Mikhaylovskiy, A.V. Kimel, Th. Rasing, J. Hellsvik and J. Lorenzana. In addition, stimulating discussions with S. Baierl, M. Barbeau, D. Bossini, U. Bovensiepen, S. Brener, E. Canovi, M. Cinchetti, E.V. Gomonay, R. Huber, S. Itin, S. Ishihara, A. Kirilyuk. J. Kroha, A. Lichtenstein, T. Oka, R.V. Pisarev, T. Satoh, S. Sayyad, H.C. Schneider, U. Staub, M. Titov, N. Tsuji and Ph. Werner are gratefully acknowledged.

This work was supported by the Nederlandse Organisatie voor Wetenschappelijk Onderzoek (NWO) by a Rubicon and a VENI grant, by the European Research Council (ERC) Advanced Grant No. 338957 (FEMTO/NANO) and Advanced Grant No. 339813 (EXCHANGE), and is part of the Shell-NWO/FOM-initiative 'Computational sciences for energy research' of Shell and Chemical Sciences, Earth and Life Sciences, Physical Sciences, FOM and STW. Part of the calculations were run on the supercomputer HLRN-II of the North-German Supercomputing Alliance.

\section*{References}
\bibliographystyle{unsrt}
\bibliography{exchangereviewbib}

\begin{thebibliography}{10}

\bibitem{beaurepaire1996}
E.~Beaurepaire, J.-C. Merle, A.~Daunois, and J.-Y. Bigot.
\newblock Ultrafast spin dynamics in ferromagnetic nickel.
\newblock {\em Phys. Rev. Lett.}, 76:4250--4253, 1996.

\bibitem{vankampen2002}
M.~van Kampen, C.~Jozsa, J.~T. Kohlhepp, P.~LeClair, L.~Lagae, W.~J.~M.
  de~Jonge, and B.~Koopmans.
\newblock All-optical probe of coherent spin waves.
\newblock {\em Phys. Rev. Lett.}, 88:227201, 2002.

\bibitem{kimel2005}
A.V. Kimel, A.~Kirilyuk, P.A. Usachev, R.V. Pisarev, A.M. Balbashov, and Th.
  Rasing.
\newblock Ultrafast non-thermal control of magnetization by instantaneous
  photomagnetic pulses.
\newblock {\em Nature}, 435:655, 2005.

\bibitem{stanciu2007}
C.D. Stanciu, F.~Hansteen, A.V. Kimel, A.~Kirilyuk, A.~Tsukamoto, A.~Itoh, and
  {Th}. Rasing.
\newblock All-optical magnetic recording with circularly polarized light.
\newblock {\em Phys. Rev. Lett.}, 99(4):047601, 2007.

\bibitem{vahaplar2009}
K.~Vahaplar, A.M. Kalashnikova, A.~V. Kimel, D.~Hinzke, U.~Nowak, R.~Chantrell,
  A.~Tsukamoto, A.~Itoh, A.~Kirilyuk, and Th. Rasing.
\newblock Ultrafast path for optical magnetization reversal via a strongly
  nonequilibrium state.
\newblock {\em Phys. Rev. Lett.}, 103:117201, 2009.

\bibitem{kirilyuk2010}
A.~Kirilyuk, A.~V. Kimel, and Th. Rasing.
\newblock Ultrafast optical manipulation of magnetic order.
\newblock {\em Rev. Mod. Phys.}, 82(3):2731--2784, 2010.

\bibitem{ostler2012}
T.A. Ostler, J.~Barker, R.F.L. Evans, R.W. Chantrell, U.~Atxitia,
  O.~Chubykalo-Fesenko, S.~El~Moussaoui, L.~Le~Guyader, E.~Mengotti, L.J.
  Heyderman, F.~Nolting, A.~Tsukamoto, A.~Itoh, D.~Afanasiev, B.A. Ivanov, A.M.
  Kalashnikova, K.~Vahaplar, J.~Mentink, A.~Kirilyuk, Th. Rasing, and A.V.
  Kimel.
\newblock Ultrafast heating as a sufficient stimulus for magnetization reversal
  in a ferrimagnet.
\newblock {\em Nat Commun}, 3:666, 2012.

\bibitem{radu2011}
I.~Radu, K.~Vahaplar, C.~Stamm, T.~Kachel, N.~Pontius, H.A. D\"urr, T.A.
  Ostler, J.~Barker, R.F.L. Evans, R.W. Chantrell, A.~Tsukamoto, A.~Itoh,
  A.~Kirilyuk, Th. Rasing, and A.~V. Kimel.
\newblock Transient ferromagnetic-like state mediating ultrafast reversal of
  antiferromagnetically coupled spins.
\newblock {\em Nature}, 472:205, 2011.

\bibitem{mathias2012}
S.~Mathias, C.~La-O-Vorakiat, P.~Grychtola, P.~Granitzkaa, E.~Turguta, J.~M.
  Shawd, R.~Adam, H.~T. Nembachd, M.~E. Siemensa, S.~Eich, C.~M. Schneider,
  T.~J. Silva, M.~Aeschlimann, M.~M. Murnane, and H.~C. Kapteyn.
\newblock Probing the timescale of the exchange interaction in a ferromagnetic
  alloy.
\newblock {\em PNAS}, 109:4792Ð4797, 2012.

\bibitem{radu2015}
I.~Radu, C.~Stamm, A.~Eschenlohr, F.~Radu, R.~Abrudan, K.~Vahaplar, T.~Kachel,
  N.~Pontius, R.~Mitzner, K.~Holldack, A.~F{\"o}hlisch, T.~A. Ostler, J.H.
  Mentink, R.F.L. Evans, R.W. Chantrell, A.~Tsukamoto, A.~Itoh, A.~Kirilyuk,
  A.V. Kimel, and Th. Rasing.
\newblock Ultrafast and distinct spin dynamics in magnetic alloys.
\newblock {\em SPIN}, 05(03):1550004, 2015.

\bibitem{lambert2014}
C-H. Lambert, S.~Mangin, B.S.D.Ch.S. Varaprasad, Y.K. Takahashi, M.~Hehn,
  M.~Cinchetti, G.~Malinowski, K.~Hono, Y.~Fainman, M.~Aeschlimann, and E.E.
  Fullerton.
\newblock All-optical control of ferromagnetic thin films and nanostructures.
\newblock {\em Science}, 345(6202):1337--1340, 2014.

\bibitem{stupakiewicz2017}
A.~Stupakiewicz, K.~Szerenos, D.~Afanasiev, A.~Kirilyuk, and Kimel~A. V.
\newblock Ultrafast nonthermal photo-magnetic recording in a transparent
  medium.
\newblock {\em Nature}, 542:71--74, 2017.

\bibitem{rhie2003}
H.-S. Rhie, H.A. D\"urr, and W.~Eberhardt.
\newblock Femtosecond electron and spin dynamics in ni/w(110) films.
\newblock {\em Phys. Rev. Lett.}, 90(24):247201, 2003.

\bibitem{carley2012}
R.~Carley et~al.
\newblock Femtosecond laser excitation drives ferromagnetic gadolinium out of
  magnetic equilibrium.
\newblock {\em Phys. Rev. Lett.}, 109:057401, 2012.

\bibitem{frietsch2015}
B.~Frietsch, J.~Bowlan, R.~Carley, M.~Teichmann, S.~Wienholdt, D.~Hinzke,
  U.~Nowak, K.~Carva, P.~M. Oppeneer, and M.~Weinelt.
\newblock Disparate ultrafast dynamics of itinerant and localized magnetic
  moments in gadolinium metal.
\newblock {\em Nat Commun}, 6:8262, 2015.

\bibitem{eiche2017}
S.~Eich, M.~Pl{\"o}tzing, M.~Rollinger, S.~Emmerich, R.~Adam, C.~Chen, H.C.
  Kapteyn, M.M. Murnane, L.~Plucinski, D.~Steil, B.~Stadtm{\"u}ller,
  M.~Cinchetti, M.~Aeschlimann, C.M. Schneider, and S.~Mathias.
\newblock Band structure evolution during the ultrafast
  ferromagnetic-paramagnetic phase transition in cobalt.
\newblock {\em Science Advances}, 3:e1602094, 2017.

\bibitem{melnikov2003}
A.~Melnikov, I.~Radu, U.~Bovensiepen, O.~Krupin, K.~Starke, E.~Matthias, and
  M.~Wolf.
\newblock Coherent optical phonons and parametrically coupled magnons induced
  by femtosecond laser excitation of the {G}d(0001) surface.
\newblock {\em Phys. Rev. Lett.}, 91:227403, 2003.

\bibitem{ju2004}
G.~Ju et~al.
\newblock Ultrafast generation of ferromagnetic order via a laser-induced phase
  transformation in ferh thin films.
\newblock {\em Phys. Rev. Lett.}, 93(19):197403, 2004.

\bibitem{thiele2004}
J.~Thiele, M.~Buess, and C.~H. Back.
\newblock Spin dynamics of the antiferromagnetic-to-ferromagnetic phase
  transition in ferh on a sub-picosecond time scale.
\newblock {\em Appl. Phys. Lett.}, 85(14):2857, 2004.

\bibitem{subkhangulov2014}
R.R. Subkhangulov, A.B. Henriques, P.H.O. Rappl, E.~Abramof, Th. Rasing, and
  A.V. Kimel.
\newblock All-optical manipulation and probing of the dÐf exchange interaction
  in eute.
\newblock {\em Sci. Rep.}, 4:4368, 2014.

\bibitem{mikhaylovskiy2015}
R.V. Mikhaylovskiy, E.~Hendry, A.~Secchi, J.H. Mentink, M.~Eckstein, A.~Wu,
  R.V. Pisarev, V.V. Kruglyak, M.I. Katsnelson, Th. Rasing, and A.V. Kimel.
\newblock Ultrafast optical modification of exchange interactions in iron
  oxides.
\newblock {\em Nat. Commun.}, 6:8190, 2015.

\bibitem{bossini2016}
D.~Bossini, S.~Dal~Conte, Y.~Hashimoto, A.~Secchi, R.V. Pisarev, Th. Rasing,
  Cerullo G., and A.~V. Kimel.
\newblock Macrospin dynamics in antiferromagnets triggered by sub-20
  femtosecond injection of nanomagnons.
\newblock {\em Nature Comm.}, 7:10645, 2016.

\bibitem{loss1998}
D.~Loss and D.~P. DiVincenzo.
\newblock Quantum computation with quantum dots.
\newblock {\em Phys. Rev. A}, 57:120--126, 1998.

\bibitem{shahbazyan2000}
T.~Shahbazyan, I.~Perakis, and M.~Raikh.
\newblock Spin correlations in nonlinear optical response: Light-induced
  {K}ondo effect.
\newblock {\em Phys. Rev. Lett.}, 84:5896--5899, 2000.

\bibitem{piermarocchi2002}
C.~Piermarocchi, Pochung Chen, L.~Sham, and D.~Steel.
\newblock Optical {RKKY} interaction between charged semiconductor quantum
  dots.
\newblock {\em Phys. Rev. Lett.}, 89:167402, 2002.

\bibitem{duan2003}
L.-M. Duan, E.~Demler, and M.~D. Lukin.
\newblock Controlling spin exchange interactions of ultracold atoms in optical
  lattices.
\newblock {\em Phys. Rev. Lett.}, 91:090402, 2003.

\bibitem{trotzky2008}
S.~Trotzky, P.~Cheinet, S.~Fšlling, M.~Feld, U.~Schnorrberger, A.M. Rey,
  A.~Polkovnikov, E.A. Demler, M.D. Lukin, and I.~Bloch.
\newblock Time-resolved observation and control of superexchange interactions
  with ultracold atoms in optical lattices.
\newblock {\em Science}, 319(5861):295--299, 2008.

\bibitem{chen2011}
Y.-A. Chen, S.~Nascimbe\`ne, M.~Aidelsburger, M.~Atala, S.~Trotzky, and
  I.~Bloch.
\newblock Controlling correlated tunneling and superexchange interactions with
  ac-driven optical lattices.
\newblock {\em Phys. Rev. Lett.}, 107:210405, 2011.

\bibitem{wall2009}
S.~Wall, D.~Prabhakaran, A.~T. Boothroyd, and A.~Cavalleri.
\newblock Ultrafast coupling between light, coherent lattice vibrations, and
  the magnetic structure of semicovalent {L}a{M}n{O}$_{3}$.
\newblock {\em Phys. Rev. Lett.}, 103:097402, 2009.

\bibitem{forst2011}
M.~F\"orst et~al.
\newblock Driving magnetic order in a manganite by ultrafast lattice
  excitation.
\newblock {\em Phys. Rev. B}, 84:241104, 2011.

\bibitem{li2013}
T.~Li et~al.
\newblock Femtosecond switching of magnetism via strongly correlated
  spin-charge quantum excitations.
\newblock {\em Nature}, 496:69--73, 2013.

\bibitem{nagaev1988}
E.L. Nagaev.
\newblock Photoinduced magnetism and conduction electrons in magnetic
  semiconductors.
\newblock {\em physica status solidi (b)}, 145(1):11--64, 1988.

\bibitem{kane1999}
B.E. Kane.
\newblock A silicon-based nuclear spin quantum computer.
\newblock {\em Nature}, 393:133--137, 1998.

\bibitem{piermarocchi2004}
C.~Piermarocchi and G.F. Quinteiro.
\newblock Coherent optical control of spin-spin interaction in doped
  semiconductors.
\newblock {\em Phys. Rev. B}, 70:235210, 2004.

\bibitem{fernandez-rossier2004}
J.~Fern{\'a}ndez-Rossier, C.~Piermarocchi, P.~Chen, A.~MacDonald, and L.~Sham.
\newblock Coherently photoinduced ferromagnetism in diluted magnetic
  semiconductors.
\newblock {\em Phys. Rev. Lett.}, 93:127201, 2004.

\bibitem{wang2007}
J.~Wang, I.~Cotoros, K.~Dani, X.~Liu, J.~Furdyna, and D.~Chemla.
\newblock Ultrafast enhancement of ferromagnetism via photoexcited holes in
  {G}a{M}n{A}s.
\newblock {\em Phys. Rev. Lett.}, 98:217401, 2007.

\bibitem{matsubara2015}
M.~Matsubara, A.~Schroer, A.~Schmehl, A.~Melville, C.~Becher,
  M.~Trujillo-Martinez, D.~G. Schlom, J.~Mannhart, J.~Kroha, and M.~Fiebig.
\newblock {U}ltrafast optical tuning of ferromagnetism via the carrier density.
\newblock {\em Nature Comm.}, 6:6724, 2015.

\bibitem{mentink2014}
J.H. Mentink and M.~Eckstein.
\newblock Ultrafast quenching of the exchange interaction in a {M}ott
  insulator.
\newblock {\em Phys. Rev. Lett.}, 113:057201, 2014.

\bibitem{mentink2015}
J.H. Mentink, K.~Balzer, and M.~Eckstein.
\newblock Ultrafast and reversible control of the exchange interaction in
  {M}ott insulators.
\newblock {\em Nat. Commun.}, 6:6708, 2015.

\bibitem{hellsvik2016}
J.~Hellsvik, J.H. Mentink, and J.~Lorenzana.
\newblock Ultrafast cooling and heating scenarios for the laser-induced phase
  transition in cuo.
\newblock {\em Phys. Rev. B}, 94:144435, 2016.

\bibitem{anderson1959}
P.~W. Anderson.
\newblock New approach to the theory of superexchange interactions.
\newblock {\em Phys. Rev.}, 115:2--13, 1959.

\bibitem{harris1967}
A.B. Harris and R.V. Lange.
\newblock Single-particle excitations in narrow energy bands.
\newblock {\em Phys. Rev.}, 157:295, 1967.

\bibitem{chao1977}
K.A. Chao, J.~Spalek, and A.M. Oles.
\newblock Kinetic exchange interaction in a narrow s-band.
\newblock {\em J. Phys. C}, 10(10):L271, 1977.

\bibitem{takahashi1977}
M.~Takahashi.
\newblock Half-filled {H}ubbard model at low temperature.
\newblock {\em J. Phys. C: Solid State Phys.}, 10(8):1289, 1977.

\bibitem{macdonald1988}
A.H. MacDonald, S.M. Girvin, and D.~Yoshioka.
\newblock ${t}/{U}$ expansion for the {H}ubbard model.
\newblock {\em Phys. Rev. B}, 37:9753--9756, 1988.

\bibitem{peierls1933}
R.~Peierls.
\newblock Zur {T}heorie des {D}iamagnetismus von {L}eitungselektronen.
\newblock {\em Z. Physik}, 80:763--791, 1933.

\bibitem{luttinger1951}
J.M. Luttinger.
\newblock The effect of a magnetic field on electrons in a periodic potential.
\newblock {\em Phys. Rev.}, 84:814--817, 1951.

\bibitem{aoki2014}
H.~Aoki, N.~Tsuji, M.~Eckstein, M.~Kollar, T.~Oka, and P.~Werner.
\newblock Nonequilibrium dynamical mean-field theory and its applications.
\newblock {\em Rev. Mod. Phys.}, 86:779--837, 2014.

\bibitem{georges1996}
A.~Georges, G.~Kotliar, W.~Krauth, and M.J. Rozenberg.
\newblock Dynamical mean-field theory of strongly correlated fermion systems
  and the limit of infinite dimensions.
\newblock {\em Rev. Mod. Phys.}, 68:13--125, 1996.

\bibitem{metzner1989}
W.~Metzner and D.~Vollhardt.
\newblock Correlated lattice fermions in $d=\infty{}$ dimensions.
\newblock {\em Phys. Rev. Lett.}, 62:324--327, 1989.

\bibitem{eckstein2010nca}
M.~Eckstein and P.~Werner.
\newblock Nonequilibrium dynamical mean-field calculations based on the
  noncrossing approximation and its generalizations.
\newblock {\em Phys. Rev. B}, 82:115115, 2010.

\bibitem{hochbruck1997}
M.~Hochbruck and C.~Lubich.
\newblock On {K}rylov subspace approximations to the matrix exponential
  operator.
\newblock {\em SIAM J. Numer. Anal.}, 34(5):1911--1925, 1997.

\bibitem{alvermann2011}
A.~Alvermann and H.~Fehske.
\newblock High-order commutator-free exponential time-propagation of driven
  quantum systems.
\newblock {\em J. Comp. Phys.}, 230(15):5930 -- 5956, 2011.

\bibitem{lichtenstein1984}
A.I. Liechtenstein, M.I. Katsnelson, and V.A. Gubanov.
\newblock {Exchange interactions and spin-wave stiffness in ferromagnetic
  metals}.
\newblock {\em J. Phys. F: Met. Phys.}, 14(7):L125, 1984.

\bibitem{lichtenstein1985}
A.I. Lichtenstein, M.I. Katsnelson, and V.A. Gubanov.
\newblock {Local spin excitations and curie temperature of iron}.
\newblock {\em Solid State Commun}, 54(4):327, 1985.

\bibitem{lichtenstein1987}
A.I. Liechtenstein, M.I. Katsnelson, V.P. Antropov, and V.A. Gubanov.
\newblock {Local spin density functional approach to the theory of exchange
  interactions in ferromagnetic metals and alloys}.
\newblock {\em J. Magn. Magn. Mater.}, 67(1):65, 1987.

\bibitem{katsnelson2000}
M.~I. Katsnelson and A.~I. Lichtenstein.
\newblock First-principles calculations of magnetic interactions in correlated
  systems.
\newblock {\em Phys. Rev. B}, 61:8906--8912, 2000.

\bibitem{secchi2013}
A.~Secchi, S.~Brener, A.~I. Lichtenstein, and M.~I. Katsnelson.
\newblock Non-equilibrium magnetic interactions in strongly correlated systems.
\newblock {\em Ann. Phys.}, 333(0):221 -- 271, 2013.

\bibitem{Floquet1883}
G.~Floquet.
\newblock {\em Ann. Sci. Ec. Normale Super.}, 12:47--88, 1883.

\bibitem{GrifoniHanggi1998}
M.~Grifoni and P.~H\"{a}nggi.
\newblock {\em Phys. Rep.}, 304:229--354, 1998.

\bibitem{itin2015}
A.P. Itin and M.I. Katsnelson.
\newblock Effective hamiltonians for rapidly driven many-body lattice systems:
  Induced exchange interactions and density-dependent hoppings.
\newblock {\em Phys. Rev. Lett.}, 115:075301, 2015.

\bibitem{bukov2016}
M.~Bukov, M.~Kolodrubetz, and A.~Polkovnikov.
\newblock Schrieffer-{W}olff transformation for periodically driven systems:
  Strongly correlated systems with artificial gauge fields.
\newblock {\em Phys. Rev. Lett.}, 116:125301, 2016.

\bibitem{kitamura2016}
S.~Kitamura and H.~Aoki.
\newblock $\ensuremath{\eta}$-pairing superfluid in periodically-driven
  fermionic {H}ubbard model with strong attraction.
\newblock {\em Phys. Rev. B}, 94:174503, 2016.

\bibitem{elliot1963}
R.J. Elliott and R.~Loudon.
\newblock The possible observation of electronic {R}aman transitions in
  crystals.
\newblock {\em Phys. Lett.}, 3(4):189 -- 191, 1963.

\bibitem{fleury1968}
P.~A. Fleury and R.~Loudon.
\newblock Scattering of light by one- and two-magnon excitations.
\newblock {\em Phys. Rev.}, 166:514, 1968.

\bibitem{shastry1990}
B.~S. Shastry and B.~I. Shraiman.
\newblock Theory of {R}aman scattering in {M}ott-{H}ubbard systems.
\newblock {\em Phys. Rev. Lett.}, 65:1068--1071, 1990.

\bibitem{lorenzana1995prb}
J.~Lorenzana and G.~A. Sawatzky.
\newblock Theory of phonon-assisted multimagnon optical absorption and bimagnon
  states in quantum antiferromagnets.
\newblock {\em Phys. Rev. B}, 52:9576--9589, 1995.

\bibitem{devereaux2007}
Thomas~P. Devereaux and Rudi Hackl.
\newblock Inelastic light scattering from correlated electrons.
\newblock {\em Rev. Mod. Phys.}, 79:175--233, 2007.

\bibitem{dalessio2014}
L.~D'Alessio and M.~Rigol.
\newblock Long-time behavior of isolated periodically driven interacting
  lattice systems.
\newblock {\em Phys. Rev. X}, 4:041048, 2014.

\bibitem{lazarides2014}
Achilleas Lazarides, Arnab Das, and Roderich Moessner.
\newblock Equilibrium states of generic quantum systems subject to periodic
  driving.
\newblock {\em Phys. Rev. E}, 90:012110, 2014.

\bibitem{zhao2004}
J.~Zhao, A.~V. Bragas, D.~J. Lockwood, and R.~Merlin.
\newblock Magnon squeezing in an antiferromagnet: Reducing the spin noise below
  the standard quantum limit.
\newblock {\em Phys. Rev. Lett.}, 93:107203, 2004.

\bibitem{zhao2006}
J.~Zhao, A.~V. Bragas, R.~Merlin, and D.~J. Lockwood.
\newblock Magnon squeezing in antiferromagnetic {M}n{F}$_2$ and {F}e{F}$_{2}$.
\newblock {\em Phys. Rev. B}, 73:184434, 2006.

\bibitem{kalashnikova2007}
A.M. Kalashnikova, A.V. Kimel, R.V. Pisarev, V.N. Gridnev, A.~Kirilyuk, and Th.
  Rasing.
\newblock Impulsive generation of coherent magnons by linearly polarized light
  in the easy-plane antiferromagnet ${\mathrm{febo}}_{3}$.
\newblock {\em Phys. Rev. Lett.}, 99:167205, 2007.

\bibitem{mikhaylovskiy2015prb}
R.V. Mikhaylovskiy, T.J. Huisman, A.I. Popov, A.K. Zvezdin, Th. Rasing, R.V.
  Pisarev, and A.V. Kimel.
\newblock Terahertz magnetization dynamics induced by femtosecond resonant
  pumping of {${\mathrm{{D}y}}^{3+}$} subsystem in the multisublattice
  antiferromagnet {${\mathrm{DyFeO}}_{3}$}.
\newblock {\em Phys. Rev. B}, 92:094437, 2015.

\bibitem{mikhaylovskiy2017}
R.~Mikhaylovskiy, T.J. Huisman, R.V. Pisarev, Th. Rasing, and A.V. Kimel.
\newblock Selective excitation of terahertz magnetic and electric dipoles in
  {${\mathrm{Er}}^{3+}$} ions by femtosecond laser pulses in
  {${\mathrm{ErFeO}}_3$}.
\newblock {\em Phys. Rev. Lett.}, 118:017205, 2017.

\bibitem{reviewyork}
R.F.L. Evans, W.J. Fan, P~Chureemart, T.A. Ostler, M.O.A. Ellis, and R.W.
  Chantrell.
\newblock Atomistic spin model simulations of magnetic nanomaterials.
\newblock {\em J. Phys.: Condens. Matter}, 26(10):103202, 2014.

\bibitem{bookuppsala}
O.~Eriksson, A~Bergman, L~Bergqvist, and J~Hellsvik.
\newblock {\em {Atomistic Spin Dynamics}}.
\newblock Oxford University Press, Oxford, 1st edition, 2017.

\bibitem{hellsvik2014}
J.~Hellsvik, M.~Balestieri, T.~Usui, A.~Stroppa, A.~Bergman, L.~Bergqvist,
  D.~Prabhakaran, O.~Eriksson, S.~Picozzi, T.~Kimura, and J.~Lorenzana.
\newblock Tuning order-by-disorder multiferroicity in {C}u{O} by doping.
\newblock {\em Phys. Rev. B}, 90:014437, 2014.

\bibitem{johnson2012}
S.~L. Johnson, R.~A. de~Souza, U.~Staub, P.~Beaud, E.~M\"ohr-Vorobeva,
  G.~Ingold, A.~Caviezel, V.~Scagnoli, W.~F. Schlotter, J.~J. Turner,
  O.~Krupin, W.-S. Lee, Y.-D. Chuang, L.~Patthey, R.~G. Moore, D.~Lu, M.~Yi,
  P.~S. Kirchmann, M.~Trigo, P.~Denes, D.~Doering, Z.~Hussain, Z.-X. Shen,
  D.~Prabhakaran, and A.~T. Boothroyd.
\newblock Femtosecond dynamics of the collinear-to-spiral antiferromagnetic
  phase transition in cuo.
\newblock {\em Phys. Rev. Lett.}, 108:037203, 2012.

\bibitem{stevens2002}
T.~E. Stevens, J.~Kuhl, and R.~Merlin.
\newblock Coherent phonon generation and the two stimulated {R}aman tensors.
\newblock {\em Phys. Rev. B}, 65:144304, 2002.

\bibitem{kittel1963}
C.~Kittel.
\newblock {\em {Quantum Theory of Solids}}.
\newblock John Wiley and Sons, New York, N.Y., 1st edition, 1963.

\bibitem{fazekas1999}
P.~Fazekas.
\newblock {\em {Lecture Notes on Electron Correlation and Magnetism}}.
\newblock World Scientific, Singapore, 1999.

\bibitem{feynman1957}
R.~P. Feynman, F.~L. Vernon~Jr., and R.~W. Hellwarth.
\newblock Geometrical representation of the schr\"odinger equation for solving
  maser problems.
\newblock {\em J. Appl. Phys.}, 28(1):49--52, 1957.

\bibitem{gridnev2008}
V.N. Gridnev.
\newblock Phenomenological theory for coherent magnon generation through
  impulsive stimulated {R}aman scattering.
\newblock {\em Phys. Rev. B}, 77:094426, 2008.

\bibitem{claassen2016}
M.~Claassen, H.~C. Jiang, B.~Moritz, and T.~P. Devereaux.
\newblock Dynamical time-reversal symmetry breaking and photo-induced chiral
  spin liquids in frustrated {M}ott insulators.
\newblock Preprint at http://arxiv.org/abs/1611.07964.

\bibitem{stepanov2017}
E.~A. Stepanov, C.~Dutreix, and M.~I. Katsnelson.
\newblock Dynamical and reversible control of topological spin textures.
\newblock {\em Phys. Rev. Lett.}, 118:157201, 2017.

\bibitem{kitamura2017}
S.~Kitamura, T.~Oka, and H.~Aoki.
\newblock Probing and controlling spin chirality in mott insulators by
  circularly polarized laser.
\newblock {\em Phys. Rev. B}, 96:014406, 2017.

\bibitem{gavrichkov2017}
V.~A. Gavrichkov, S.~I. Polukeev, and S.`~G. Ovchinnikov.
\newblock Contribution from optically excited many-electron states to the
  superexchange interaction in {M}ott-{H}ubbard insulators.
\newblock {\em Phys. Rev. B}, 95:144424, 2017.

\bibitem{eckstein2017}
M.~Eckstein, J.H. Mentink, and P.~Werner.
\newblock Designing spin and orbital exchange hamiltonians with ultrashort
  electric field transients.
\newblock Preprint at http://arxiv.org/abs/1703.03269.

\bibitem{sato2016}
M.~Sato, S.~Takayoshi, and T.~Oka.
\newblock Laser-driven multiferroics and ultrafast spin current generation.
\newblock {\em Phys. Rev. Lett.}, 117:147202, 2016.

\bibitem{meyer2017}
U.~Meyer, G.~Haack, C.~Groth, and X.~Waintal.
\newblock Control of the oscillatory interlayer exchange interaction with
  terahertz radiation.
\newblock {\em Phys. Rev. Lett.}, 118:097701, 2017.

\bibitem{kampfrath2010}
T.~Kampfrath, A.~Sell, G.~Klatt, A.~Pashkin, S.~M\"ahrlein, T.~Dekorsy,
  M.~Wolf, M.~Fiebig, A.~Leitenstorfer, and R.~Huber.
\newblock Coherent terahertz control of antiferromagnetic spin waves.
\newblock {\em Nat. Phot.}, 5:31, 2010.

\bibitem{baierl2016a}
S.~Baierl, M.~Hohenleutner, T.~Kampfrath, A.K. Zvezdin, A.V. Kimel, R.~Huber,
  and R.V. Mikhaylovskiy.
\newblock Nonlinear spin control by terahertz-driven anisotropy fields.
\newblock {\em Nat. Phot.}, 10:715--718, 2016.

\bibitem{baierl2016b}
S.~Baierl, J.H. Mentink, M.~Hohenleutner, L.~Braun, T.-M. Do, C.~Lange,
  A.~Sell, M.~Fiebig, G.~Woltersdorf, T.~Kampfrath, and R.~Huber.
\newblock Terahertz-driven nonlinear spin response of antiferromagnetic nickel
  oxide.
\newblock {\em Phys. Rev. Lett.}, 117:197201, 2016.

\bibitem{bruck2007}
E.~Br{\"u}ck, O.~Tegus, D.T.C. Thanh, and K.H.J. Buschow.
\newblock Magnetocaloric refrigeration near room temperature (invited).
\newblock {\em Journal of Magnetism and Magnetic Materials}, 310(2):2793 --
  2799, 2007.
\newblock Proceedings of the 17th International Conference on Magnetism.

\end{thebibliography}

\end{document}